\documentclass[tradiabstract]{aa}
\usepackage[dvips,final]{graphicx}
\usepackage{txfonts}
\usepackage{natbib}
\usepackage{multirow}
\usepackage{rotating}
\bibpunct{(}{)}{;}{a}{}{,}

\begin{document}
   \title{Mid\,-\,infrared interferometry of massive young stellar objects II}
   \subtitle{Evidence for a circumstellar disk surrounding the Kleinmann\,-\,Wright object
   \thanks{Based on observations within the ESO programs 075.C-0755(AB), 077.C-0440(ABC).}}
   \author{R.~Follert\inst{1,2}\fnmsep\thanks{Fellow of the International Max Planck Research School for Astronomy and Cosmic Physics at the University of Heidelberg (IMPRS-HD)}
          \and H.~Linz\inst{1}
          \and B.~Stecklum\inst{2}
	  \and R.~van Boekel\inst{1}
          \and Th.~Henning\inst{1}
	  \and M.~Feldt\inst{1}
          \and T.M.~Herbst\inst{1,3}
	  \and Ch.~Leinert\inst{1}
}
   \offprints{R.~Follert\inst{1}}
   \institute{Max-Planck-Institut f\"ur Astronomie, 
              K\"onigstuhl 17, D-69117 Heidelberg, Germany\\
              \email{[follert,linz,boekel]@mpia-hd.mpg.de}
	 \and Th\"uringer Landessternwarte Tautenburg, Sternwarte 5,
	      D--07778 Tautenburg, Germany
	      \email{stecklum@tls-tautenburg.de}
	 \and Herzberg Institute of Astrophysics, National Research Council of Canada, 5071 West Saanich Rd, Victoria, BC, V9E 2E7 Canada
}
   \date{Received May 30, 2008 }


\newcommand{\degree}{\ensuremath{^\circ}}

\abstract{The formation scenario for massive stars is still under discussion. To further constrain current theories, it is vital to spatially resolve the structures from which material accretes onto massive young stellar objects (MYSOs). Due to the small angular extent of MYSOs, one needs to overcome the limitations of conventional thermal infrared imaging, regarding spatial resolution, in order to get observational access to the inner structure of these objects. We employed mid\,-\,infrared interferometry, using the MIDI instrument on the ESO\,/\,VLTI, to investigate the Kleinmann\,-\,Wright Object, a massive young stellar object previously identified as a Herbig Be star precursor. Dispersed visibility curves in the N\,-\,band (8\,--\,13 $\mu$m) have been obtained at 5 interferometric baselines. We show that the mid\,-\,infrared emission region is resolved. A qualitative analysis of the data indicates a non\,-\,rotationally symmetric structure, e.g. the projection of an inclined disk. We employed extensive radiative transfer simulations based on spectral energy distribution fitting. Since SED\,-\,only fitting usually yields degenerate results, we first employed a statistical analysis of the parameters provided by the radiative transfer models. In addition, we compared the ten best\,-\,fitting self\,-\,consistent models to the interferometric observations. Our analysis of the Kleinmann\,-\,Wright Object suggests the existence of a circumstellar disk of 0.1\,M$_\odot$ at an intermediate inclination of 76\degree, while an additional dusty envelope is not necessary for fitting the data. Furthermore, we demonstrate that the combination of IR interferometry with radiative transfer simulations has the potential to resolve ambiguities arising from the analysis of spectral energy distributions alone.}

\maketitle
%

\section{Introduction}
One of the central unsolved issues in the study of massive star formation is to which extent the accretion mechanism is comparable to the better\,-\,understood processes in low\,-\,mass star formation, e.g., accretion of material from a disk structure by the forming star at certain evolutionary stages. Although there are several theoretical scenarios for massive star formation \citep{2007ARA&A..45..481Z}, the lack of observational data on the inner structure of massive young stellar objects (MYSOs)  prevents a deeper understanding of massive star formation. There are two main reasons for this problem: first, these objects are usually deeply embedded, so observations need to be done in the mid\,-\,infrared (MIR) or at even longer wavelengths. Second the angular extent of these objects is rather small. In order to resolve the inner structures of YSOs in general, high angular resolution observations are required. Since diffraction\,-\,limited observations, even with 8\,m-class telescopes, result in resolvable angular scales of a few hundred milliarcseconds, interferometric techniques need to be applied to assess the innermost structures. On the other hand, analysis of the data obtained by interferometric techniques, specifically from the dispersed visibility curves, is far from trivial. This holds in particular when the number of baselines is small. In this respect, Monte Carlo radiative transfer (MCRT) simulations represent an important tool to analyze visibility curves, by calculating synthetic curves for comparison with the modeled to the 
observed data.\vspace{5pt}\\ 
We are currently conducting a large interferometric study of MYSOs with the Very Large Telescope Interferometer (VLTI) \citep{2008JPhCS.131a2024L}. Most of the targets were taken from a list of BN\,-\,type objects \citep{1984AN....305...67H}. First results on the object M8E\,-\,IR have been reported by \citet{2009A&A...505..655L}. In the present paper, we concentrate on the Kleinmann\,-\,Wright Object (\citealt{1973ApJ...185L.131K} - also known as M17 SW IRS 1, hereafter KWO). Based on NIR spectroscopic data, \citet{1998A&A...332..999P}  classified the KWO as a transition object between a deeply embedded MYSO and a Herbig Be star. \citet{2004A&A...427..849C} presented for the first time iJHK images, in which the KWO is clearly resolved into two separate sources. They determined a projected separation of 1\farcs2, or 2650 AU, assuming a distance of 2.2 kpc. \citet{2004A&A...427..849C} assumed that the second component (No. 2 in their paper) contributes a significant amount of flux only in the visual and near infrared wavelength regime. Here, we focus our research on the brighter component (No. 1 in \citealt{2004A&A...427..849C}). Furthermore, Chini et al. used multi\,-\,color photometric data, ranging from the Gunn i band to the MIR, and a TIMMI2 spectrum to fit a simple radiative transfer (RT) model, assuming a spherical dust distribution. Due to excess in the infrared part of the spectral energy distribution (SED), \citet{2004A&A...427..849C} proposed that the KWO is a heavily embedded Herbig Be star, in agreement with the work of \citet{1998A&A...332..999P}. In this paper, we present new MIR interferometry for the KWO, and we re\,-\,evaluate the SED for this source, based now on 2D RT simulations. We discuss the statistics of the model parameters, and how the interferometric results constrain these parameters.
\section{Observations and data reduction}
\begin{figure}[ht!]
\centering
\includegraphics[width=0.9\linewidth,keepaspectratio]{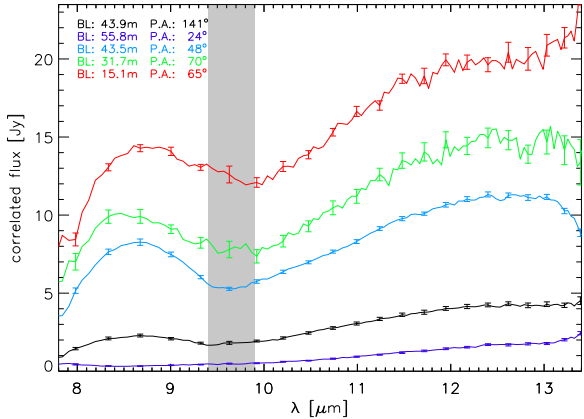}
\caption{\small The measured correlated fluxes. \label{pic:corrflux}}
\end{figure}

The interferometric data were obtained with the MIDI instrument installed at the VLTI \citep{2003SPIE.4838..893L}. The KWO was observed during five nights at five different projected baseline\,/\,projected angle configurations in the N\,-\,Band (8\,-\,13\,$\mu$m). For the first three nights of observation (2005-06-24, 2005-06-26, 2006-05-18), the 8.2\,m diameter Unit Telescopes (UTs) were used. For the other two nights, the 1.8\,m Auxiliary Telescopes (ATs) were used for the observation. At the time of submission, this paper represented the first published analysis of MYSO data using the ATs.
{\begin{table}[ht!]
\caption{\small List of observed objects. All objects other than the KWO are calibrator stars (see above). The objects marked with an asterisk are included in the subsequent analysis. See text for details. The third column, labelled B, lists the projected baseline in meters. \label{tab:MIDI-obs}}
\tiny
\begin{tabular}{|c|c|c|c|c|}\hline
\multirow{2}{*}{Object}&\multicolumn{1}{|c}{Date and }&\multicolumn{1}{|c}{B}&\multicolumn{1}{|c|}{PA [\degree]}& \multirow{2}{*}{UTs}\\
&\multicolumn{1}{|c}{Time}&\multicolumn{1}{|c}{[m]}&\multicolumn{1}{|c|}{E of N}& \\\hline
HD107446*&2005-06-24T03:24:09&61.51&167.2&3\,-\,4\\ 
HD168415*&2005-06-24T04:22:42&62.34&108.6&3\,-\,4\\
HD169916&2005-06-24T07:04:18&56.89&127.4&3\,-\,4\\ 
HD169916&2005-06-24T07:59:14&52.06&138.1&3\,-\,4\\ 
KWO*&2005-06-24T08:23:34&43.87&140.7&3\,-\,4\\
HD169916*&2005-06-24T08:48:54&47.75&150.3&3\,-\,4\\
HD177716&2005-06-24T09:59:36&47.07&160.2&3\,-\,4\\ \hline
HD169916*&2005-06-26T00:49:52&55.84&-6.5&1\,-\,2\\ 
HD107446&2005-06-26T01:44:57&38.57&57.5&1\,-\,2\\
HD169916&2005-06-26T02:29:52&55.88&8.0&1\,-\,2\\
HD145544&2005-06-26T03:58:14&43.41&44.6&1\,-\,2\\
KWO*&2005-06-26T04:25:11&55.79&24.3&1\,-\,2\\\hline
HD95272*&2006-05-18T00:36:17&46.63&43.4&2\,-\,3\\
HD169916&2006-05-18T04:07:17&43.99&10.4&2\,-\,3\\
HD101666&2006-05-18T05:41:02&26.03&58.0&2\,-\,3\\
HD120404&2006-05-18T06:42:24&28.96&100.1&2\,-\,3\\
HD120404&2006-05-18T07:36:18&25.98&114.7&2\,-\,3\\
HD168415&2006-05-18T09:27:57&44.99&47.8&2\,-\,3\\
KWO*&2006-05-18T09:59:02&43.52&48.4&2\,-\,3\\
HD169916*&2006-05-18T10:28:38&40.01&52.2&2\,-\,3\\\hline
&&&&ATs\\\hline
HD150798      &2006-08-09T23:30:24 &31.6   &65.6&D0\,-\,G0\\
KWO*           &2006-08-10T01:27:29 &31.7   &69.6&D0\,-\,G0\\
HD211416       &2006-08-10T01:52:18 &31.5   &22.6&D0\,-\,G0\\
HD168454*       &2006-08-10T02:12:13 &31.8   &73.9&D0\,-\,G0\\\hline
KWO*            &2006-08-13T00:22:43 &15.1  &64.76&E0\,-\,G0\\
HD211416       &2006-08-13T00:48:05 &15.7  &  9.9&E0\,-\,G0\\
HD150798       &2006-08-13T01:40:00 &14.9  &-81.7&E0\,-\,G0\\
HD150798       &2006-08-13T02:46:00 &14.2  &-65.5&E0\,-\,G0\\
HD150798       &2006-08-13T03:42:34 &13.6  &-50.7&E0\,-\,G0\\
HD206778*      &2006-08-13T04:35:19 &15.4 &74.8&E0\,-\,G0\\\hline
\end{tabular}
\end{table}}\\
Table \ref{tab:MIDI-obs} lists the observations, dates, times, telescopes used, as well as the associated baseline lengths and orientations. The observations followed the procedure described in \citet{2004A&A...423..537L}. MIDI was set to work in HighSens\,-\,mode. The beams from the two telescopes were first combined to determine the interferometric signal or correlated flux. While measuring the correlated flux, the optical path difference (OPD) is first changed on a coarser scale (some millimeters) to search for the interferometric fringes. Once they are found, the OPD is swept continuously around the zero OPD (the white light fringe) with an amplitude of approximately 80\,$\mu$m at a frequency of a few hundred scans in three to four minutes. 
After the correlated flux is determined, the photometric signal (or uncorrelated flux) is measured.\vspace{5pt}\\
For that purpose, the light from one telescope is blocked, allowing measurement of the flux from the other and vice versa. During the recording of the photometric signal, chopping with a throw of 10$''$ and a frequency of roughly 2 Hz is applied to remove the background signal. Appropriate calibrator stars were observed in the same way immediately before and after the KWO (see table \ref{tab:MIDI-obs} for details), in order to account for instrumental effects which distort the visibility curve.\vspace{5pt}\\
The MIDI prism was used as the dispersing element, providing a spectral resolution of $R\approx30$. The data obtained with MIDI were first reduced using the MIA+EWS package (V1.5.1), developed at the MPIA Heidelberg and the Leiden Observatory (see {\tt http://www.mpia-hd.mpg.de/MIDI/} or \citet{2004SPIE.5491..715J}).
\begin{figure}[ht!]
\centering
\includegraphics[width=0.9\linewidth,keepaspectratio]{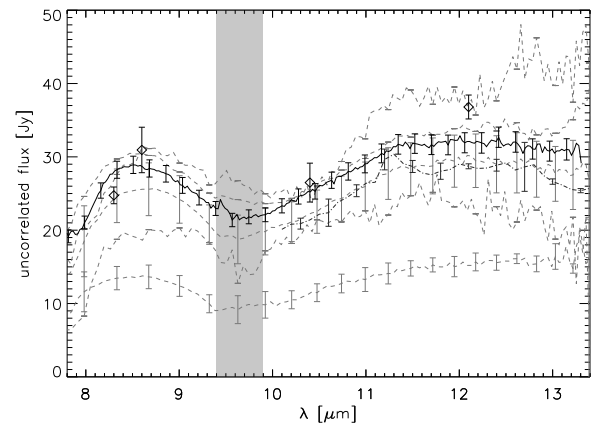}
\caption{\small Photometry of the KWO. The dashed, gray lines depict the spectra as obtained by MIDI, the dash\,-\,dotted, dark gray line archival IRS data, the diamonds the N\,-\,band SED from \citet{2004A&A...427..849C}, and the bold, black line the spectrum used for further analysis. The vertical gray bar denotes the wavelength regime of the atmospheric ozone absorption feature. \label{pic:spectra}}
\end{figure}\\
We found that the photometry results, when calibrating with different calibrators belonging to one night of observation, were not consistent (see Fig. \ref{pic:spectra}). This is a clear sign that the nights of observation were far from photometric. Hence, we employed a different data reduction scheme. First, the correlated fluxes of the KWO for each night of observation were determined and afterward calibrated by the corresponding set of correlated fluxes of the calibrator stars (see Fig. \ref{pic:corrflux}). The correlated flux is less affected by atmospheric emission, due to the differential recording technique. For determining the correlated flux, two subsequent 180\degree \space phase\,-\,shifted measurements are subtracted from each other. Due to the high fringe scan speed, the background is subtracted far more efficiently than by the chopping procedure. Accordingly, the variations in the correlated fluxes were negligible within the errors. The only exception are artifacts caused by the atmospheric absorption features (e.g. the ozone feature at 9.4\,-\,9.9\,$\mu$m). If the science target and the calibrator are observed at different air masses, the calibrator will be more (or less) affected by the atmospheric absorption, which may result in emission\,-\,(absorption\,-) like artifacts in the calibrated correlated fluxes. Hence, only calibrators observed at air masses comparable to those of the KWO were taken into account for the subsequent analysis (marked with an asterisk in table \ref{tab:MIDI-obs}). Nevertheless, the correlated fluxes in the wavelength range covering the ozone feature might still be slightly affected; hence, we omitted these data from the analysis.
\\
In order to deduce the calibrated visibilities, the calibrated correlated fluxes need to be divided by the calibrated uncorrelated flux (i.e. the spectrum).\vspace{5pt}\\
Since we wanted to avoid the non\,-\,photometric data obtained by MIDI, we instead used TIMMI2 data (see Fig. \ref{pic:spectra}, solid bold black line). The ESO archive contains mid-infrared grism spectroscopy of the KWO between 8.0 and 13.0  $\mu{\rm m}$ obtained with TIMMI2 \citep{2000SPIE.4008.1132R} with a nominal spectral resolution of 230. The observations were performed within the ESO programme 71.C-0185(A) on July 25, 2003. Using a slit width of 1\farcs2, the measurements were done with chopping and nodding throws of 10$''$, with a total integration time of seconds. The star HD169916 served as spectrophotometric standard star.  For the re\,-\,reduction of these data, we used the TIMMI2 pipeline by \citet{2004A&A...414..123S}. The same spectrum was taken into account for the subsequent SED fitting. For that purpose, the data were re\,-\,sampled to match the wavelength grid used by the Online SED model fitter (\citet{2007ApJS..169..328R}, see also section \ref{subsec:SEDFIT}).\vspace{5pt}\\
Furthermore, Spitzer IRS data from the archive and the SED as specified in \citet{2004A&A...427..849C} complemented the spectral information. The post\,-\,basic calibrated IRS data (Request Key 11546624, PI M.\,Wolfire) was reduced with the standard software version 15.3.0 and subsequently re\,-\,binned. The KWO is quite luminous in the NIR. Thus, the Spitzer IRAC frames are overexposed. There are no Spitzer MIPS data of the KWO.  Unfortunately, no (sub-)\,millimeter continuum data are available for the KWO. The SED used in the subsequent analysis is summarized in table \ref{tab:SED-Data}. For the error analysis, variations in the correlated flux (during one measurement cycle) as specified by the data reduction tool were taken into account, as well as the error specified for the TIMMI2 spectrum. If more than one calibrator provided usable data, the respective visibility curves were finally averaged. \begin{figure}[ht!]
\centering
\includegraphics[width=0.8\linewidth,keepaspectratio]{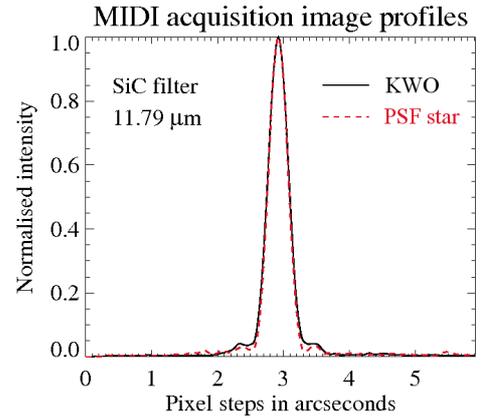}
\caption{\small Comparison between radial cuts of image profiles of the KWO and the calibrator star HD169916, derived from MIDI UT acquisition images. \label{pic:PSF}}
\end{figure}\\
In addition to the interferometric data, we also took acquisition images from the UT measurements at 11.8\,$\mu$m. After background subtraction, we compared the profiles of the science target and a calibrator star. They are equal within the errors, which is remarkable since the observation of the calibrator has a clearly better adaptive optics correction (on\,-\,source guiding). Hence the KWO is not resolved in the MIR with single 8\,m\,-\,class telescopes (see Fig. \ref{pic:PSF}).\\
{\centering
\begin{table}[ht!]
\caption{\small Flux densities of the KWO \label{tab:SED-Data}}
\begin{tabular}{|c|c|c|c|c|}\hline
\multicolumn{1}{|c}{wavelength}&\multicolumn{1}{|c}{F$_\nu$}&\multicolumn{1}{|c}{$\sigma_{{\mathrm F}_\nu}$}&\multicolumn{1}{|c}{aperture}& \multicolumn{1}{|c|}{data}\\
\multicolumn{1}{|c}{[$\mu$m]}&\multicolumn{1}{|c}{[mJy]}&\multicolumn{1}{|c}{[mJy]}&\multicolumn{1}{|c}{arcsec}& \multicolumn{1}{|c|}{source}\\\hline
0.8 & 0.025&0.005&2&  Chini et al. (2004)\\
1.25& 7    &1    &2&  Chini et al. (2004)\\
1.65& 91   &9    &2&  Chini et al. (2004)\\
2.2 & 550  &60   &2&  Chini et al. (2004)\\
3.9 & 4400 &1000 &2.5&  Chini et al. (2004) \\
4.4 & 9700 & 4500 &18 & MSX B (archive)\\
4.6 & 5800 &1500 &3& Chini et al. (2004) \\
8.06&22920&930  &1.2&TIMMI2\\
8.50&28890&1180 &1.2&TIMMI2\\
8.97&26670&1090 &1.2&TIMMI2\\
9.44&24110&1030 &1.2&TIMMI2\\
9.95&22310&940  &1.2&TIMMI2\\
10.49&25920&1080&1.2&TIMMI2\\
11.05&29510&1240&1.2&TIMMI2\\
11.65&31450&1320&1.2&TIMMI2\\
12.27&31730&1380&1.2&TIMMI2\\
12.93&31090&1370&1.2&TIMMI2\\
14.7&33070 &1650 &18& Chini et al. (2004)\\ 
17.8&55488 &13900& 5& Chini et al. (2004)\\
21.3&68760 &3440 &18& Chini et al. (2004)\\
23.04&105000&6100&12& IRS\\
26.96&129000&5900&12& IRS\\
31.56&156000&7100&12& IRS\\\hline
\end{tabular}
\end{table}}\\
\section{Results}
The analysis of spectrally dispersed visibility curves is non\,-\,trivial. Analytical solutions can only be found if one assumes very simple models and/or geometrical shapes. For this reason, we first fit the SED to pre\,-\,calculated models. The set of parameters which describe the respective model is then used as input for a Monte Carlo radiative transfer model. The resulting synthetic images are finally Fourier\,-\,transformed in order to produce synthetic visibility curves which can be compared to the measurements \citep{2009A&A...505..655L}.
\subsection{SED fitting}\label{subsec:SEDFIT}
The SED data served as input for the Online SED model fitter\footnote{\tt
 http://caravan.astro.wisc.edu/protostars/index.php} \citep{2007ApJS..169..328R}. The fitter is based on a grid of 200,000 pre\,-\,calculated, 2D Monte Carlo radiative transfer models computed with the code of \citet{2003ApJ...598.1079W}. Each of these models is characterized by a set of 28 parameters, which are not all independent of each other (for details see \citet{2006ApJS..167..256R} and \citet{2008ASPC..387..290R}). The models include the properties of the central object (mass, radius, temperature), the envelope (accretion rate, outer radius, inner cavity density and opening angle), the disk (mass, inner and outer radius, accretion rate, scale height factor, flaring angle) and the ambient density. These parameters have been varied to cover a large range of possible configurations of young stellar objects, yet the parameter volume is not covered uniformly. The result is a first grid\,-\,like set of 20,000 models. For each of these models, the SED was calculated for 10 different inclination angles (defined between the axis of symmetry of the model and the line of sight), distributed equally over the cosines of the inclination angle. Hence, the final grid consists of 200,000 models. The grain model used consists of a mixture of astronomical silicates and graphite in solar abundance, where the optical constants are taken from \citet{1993ApJ...402..441L}. In addition, the optical properties are averaged over the size distribution \citep[taken from][]{1994ApJ...422..164K} and composition. For further details see \citet{2006ApJS..167..256R}.\vspace{5pt}\\
In addition to the SED, the user is required to specify some assumptions on the distance and interstellar extinction toward the object as input for the Online SED model fitter. These assumptions are treated as boundary values in the fitting procedure. The fitter then performs a global $\chi ^2$ minimization procedure to evaluate which parameter combination fits the observed SED best. The results are presented ordered by increasing deviation between the modeled and specified SEDs. For each of these models, a parameter file is available which contains the set of parameters describing this particular model. These files then serve as input to the Monte Carlo radiative transfer code.\vspace{5pt}\\
For this work, the distance and the interstellar extinction were chosen in accordance with \citet{2004A&A...427..849C}, who proposed a distance d\,=\,2.2\,kpc and a total visible extinction of $A_{v}\approx24$\,mag, including extinction due to both the intervening interstellar matter and the local circumstellar structures. The recent re\,-\,determination by \citet{2008ApJ...686..310H}, who found d\,=\,2.1\,$\pm$\,0.2\,kpc for the M17 SW star cluster, was also taken into account. Hence, we adopted as boundaries for the fitting procedure d\,=\,1.9\,-\,2.3\,kpc, while the interstellar extinction was assumed to be larger than 5 mag.We want to point out again that the distance, the interstellar as well as the total extinction are treated as free parameters (within the boundaries). Figure \ref{pic:SEDFit} presents the ten best SED fits.\vspace{5pt}\\
The Spitzer IRS data were considered as upper limits. There will be a significant contribution to the luminosity by the interstellar material surrounding the KWO, due to the  much larger aperture of IRS at these wavelengths. Apparently, the fitted curve overestimates the flux at 3.9 and 4.6\,$\mu$m. In order to assess the result, we ran the SED fitter again, including each data point shortward of 8\,$\mu$m twice. The resulting set of solutions indeed involves some spectra which agree on the relatively low flux at 3.9 and 4.6\,$\mu$m. Yet, we found that all these solutions cannot account for the pronounced silicate feature (see Fig. \ref{pic:spectra}). In addition, we included a MSX B2 data point from the archive. Due to the uncertain calibration of the MSX B2 data, the associated error is large; hence, it does not have a strong effect on the SED fit. Nevertheless the MSX B2 flux and the best fit match quite well. Hence, we kept the result as presented in Fig. \ref{pic:SEDFit}.
{\begin{figure}[!ht]
\centering
\includegraphics[width=\linewidth,keepaspectratio]{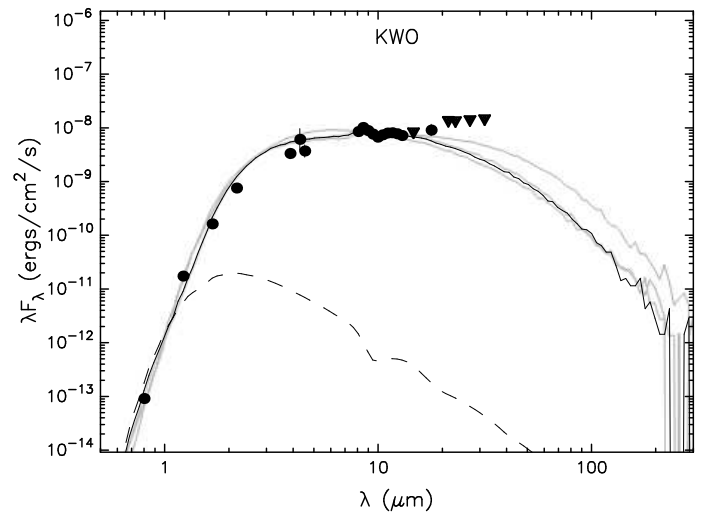}
\caption{\small Best SED fits, obtained with the Online SED model fitter. The black solid line corresponds to the best fitting model, the gray lines to the 9 next best fits. The dashed line shows the stellar photosphere model with the best\,-\,fit interstellar extinction taken into account. The dots and triangles represent the data the fit was applied to, dots denote regular data points and triangles upper limits, respectively.}
\label{pic:SEDFit}
\end{figure}}\\
The compiled SED used for this work is based on data covering the peak and Wien side of the SED. No published data exist for the Rayleigh-Jeans part, which could distinguish KWO from the surrounding high level of extended (sub-)\,millimeter flux present in the  M17 SW region. Therefore, the SED of KWO is not constrained at far infrared and submm wavelengths, and several model configurations fit the observed optical\,/\,infrared SED almost equally well. For this reason, a statistical approach to constrain the meaningful parameter space was adopted (see Sec. \ref{sect:Discussion}). We created histograms of the different model parameters (accounting for the 1000 best fitting models), with each count weighted by $1/\chi^2$. High densities in the respective parameter space hence give an indication of the probability that a certain parameter will give a value in the associated range.
\subsection{Radiative transfer modeling and synthetic visibilities}\label{sec:RTmodel}
\begin{figure}[!ht]
\centering
\includegraphics[width=0.65\linewidth,keepaspectratio]{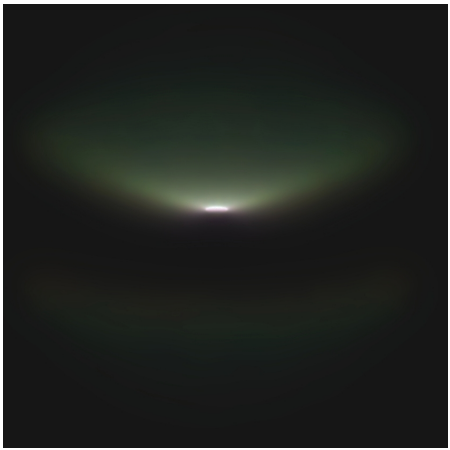}
\caption{\small Synthetic RGB composite image for model 3003929 (see Table \ref{longtable2}) at an inclination angle of 75$^\circ$. The 12.5\,$\mu$m bin is depicted in red, the 10.5\,$\mu$m bin in green and the 8.5\,$\mu$m bin in blue. The total visible area is 1200$\times$1200\,AU. \label{pic:model}}
\end{figure}

For the 10 best\,-\,fitting configurations, MCRT simulations were performed using the {\sc ttsre} code of \citet{2003ApJ...598.1079W}. The simulation was run each time using a total of 100 million test photons. We modified the MCRT code for our purposes by defining 11 narrow filters centered at equally spaced wavelengths covering the N\,-\,band. Hence, we obtained for each model 11 synthetic images. Figure \ref{pic:model} shows an example of the frames created by this procedure. We combined 3 synthetic images of different wavelength to create an RGB composite. In addition, we applied a logarithmic intensity scaling in order to reveal the diffuse outer regions of the structure. The picture was adaptively smoothed to reduce the graininess inherent to Monte Carlo simulations.\vspace{5pt}\\
Each of these synthetic images was first embedded in a larger, empty frame, since the MCRT code only allows for a certain spatial resolution in combination with a respective maximal spatial outer scale. Then, each image was Fourier-transformed in order to obtain the related (u\,,\,v)\,-\,spatial frequency spectrum. Due to the increased spatial outer scale, the resolution in Fourier\,-\,space was refined, which in turn helped to avoid numerical discretization effects.
\\
Next, the Cartesian coordinates of the (u\,,\,v)\,-\,plane were transformed to polar coordinates. The synthetic visibility curves were constructed by adopting the values from the synthetic (u\,,\,v)\,-\,spectra from each of the respective wavelength bin images at the 
spatial frequencies corresponding to the projected baselines used in the observations. There is one remaining degree of freedom: while the inclination angle \textit{i} is fixed due to the SED fitting and the respective MCRT modeling, the symmetry axis of the synthetic images runs north-south, which is not necessarily true for the science target. Hence, an additional angular parameter is required to account for the observed position angle. This parameter was derived by minimizing the deviation between the synthetic visibility curves and the measured visibility curves by rotating the u\,-\,v plane. Hence, we obtained for each model spectrally dispersed, synthetic visibility curves at the projected baselines used for the observation. These will be compared to the measured visibilities in \ref{sec:comparisson}.\vspace*{5pt}\\
This approach is similar to \citet{2010A&A...515A..45D}, who treated the MYSO W33A. They use the RT models of \citet{2003ApJ...598.1079W} to generate synthetic visibility curves as well. Yet, \citet{2010A&A...515A..45D} were able to constrain various model parameters in advance due to better availability of data for W33A, especially in the submm wavelength range.
\subsection{Interferometric observations}
\begin{figure}[ht!]
\centering
\includegraphics[width=0.9\linewidth,keepaspectratio]{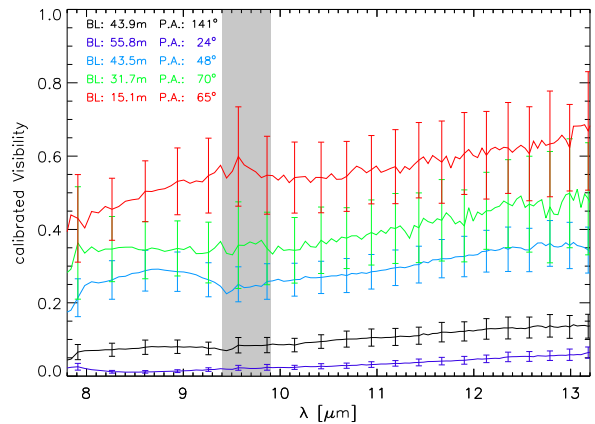}
\caption{\small Calibrated visibility curves.\label{pic:vis-curves}}
\end{figure}

The measured calibrated dispersed visibilities appear in Fig. \ref{pic:vis-curves}. The depicted errors represent 3$\sigma$. The visibility amplitudes are smaller than unity for all observations. Hence, the KWO is clearly resolved on all baselines. Moreover, the silicate absorption feature we see in the total intensity spectra is not seen in any of the visibility curves. This is due to the fact that the feature is equally strong in both the correlated and uncorrelated flux. Since we see the same effect in every visibility curve, although they are measured on clearly distinct baseline lengths, this indicates that the absorbing dust is situated in foreground structures not directly associated with the KWO. The existence of such a dust screen in front of the KWO was already proposed by \cite{2004A&A...427..849C} and \cite{1998A&A...332..999P}, who suggest that the KWO might actually be located behind the SW obscuration of M17.
\section{Discussion}\label{sect:Discussion}
\subsection{Statistical reduction of the parameter volume}\label{subsec:stat}
SED data fitting of MYSOs is in general ambiguous. In order to investigate how the observational errors translate into variations of the RT model parameters, we followed the approach of \citet{2008arXiv0808.0619P}. Note that the histograms (see section \ref{subsec:SEDFIT}) represent the distribution of the respective parameters taken from the set of the 1000 best\,-\,fitting models as  obtained from the Online SED model fitter, no a priori assumptions were involved.
\subsubsection{The stellar properties}
\begin{figure}[!ht]
\centering
\includegraphics[width=0.8\linewidth,keepaspectratio]{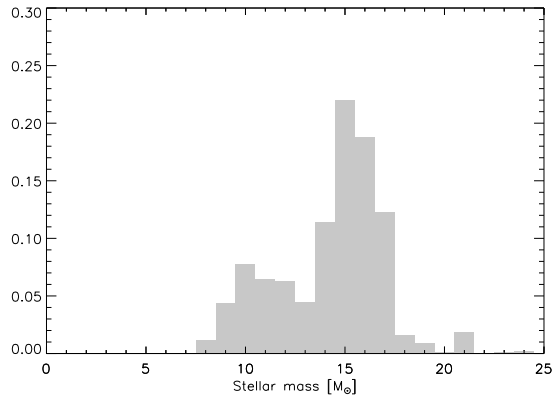}
\caption{\small Distribution of the stellar mass [M$_{\odot}$]}
\label{fig:Starmass}
\end{figure}

\begin{figure}[!ht]
\centering
\includegraphics[width=0.8\linewidth,keepaspectratio]{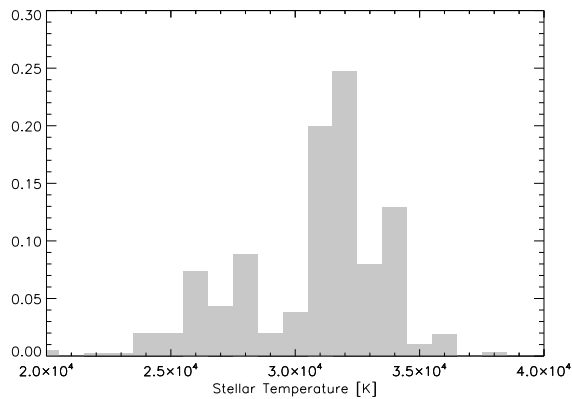}
\caption{\small Distribution of the stellar temperature [K] \label{fig:Startemp}}
\end{figure}

Figures \ref{fig:Starmass} and \ref{fig:Startemp} show the distribution of the fitted stellar masses and temperatures, respectively. The fitted stellar temperature is strongly peaked around 32000 K. The stellar mass distribution is peaked around 15\,M$_\odot$, with a side lobe towards 10\,M$_\odot$. Hence, the most probable embedded star for the KWO would be an early B star. 
\subsubsection{Disk and envelope}
\begin{figure}[!ht]
\centering
\includegraphics[width=0.8\linewidth,keepaspectratio]{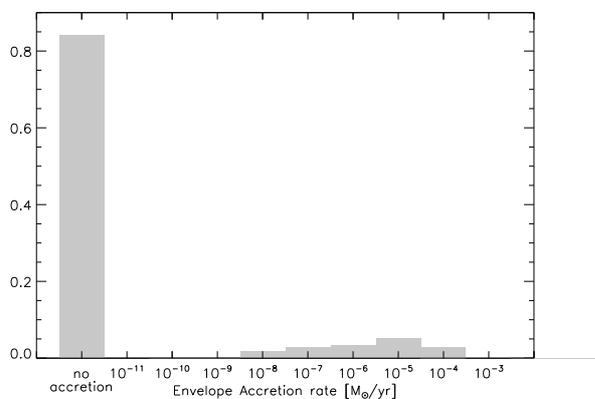}
\caption{\small  Distribution of the envelope accretion rate [$\frac{M_\odot}{yr}$]\label{fig:EAccr}}
\end{figure}

Figure \ref{fig:EAccr} plots the envelope accretion rate. It is interesting to note that most of the models fitting the SED have no envelope at all: these models consist of the central star and a disk only. In the \citet{2007ApJS..169..328R} models, this is denoted by setting the envelope accretion rate formally to zero. This is in contrast to \citet{2004A&A...427..849C} who fit the SED with a spherical dust distribution with a radial density profile varying as r$^{-0.5}$, which might be interpreted as an envelope\,-\,like structure. Due to the inability of this approach to properly reproduce the silicate feature, one of their suggestions was a disk\,-\,like structure, which would produce silicate features both in emission and absorption, depending on the disk orientation. \citet{2004A&A...427..849C} identify the reflection nebula surrounding the KWO with this spherical dust distribution. The arguments made above and our interferometric measurements (see sections \ref{sec:Gen_Visis} and \ref{sec:comparisson}) favor a configuration without a small\,-\,scale optically thick dust shell directly enveloping the KWO. Thus, the scattering material in the reflection nebula is probably not bound to the KWO itself. Instead, we suggest that the reflecting material is either a remnant of the natal molecular cloud or it is situated in front of the KWO along the line of sight (as already proposed by \citet{1998A&A...332..999P}.
\begin{figure}[ht!]
\centering
\includegraphics[width=0.8\linewidth,keepaspectratio]{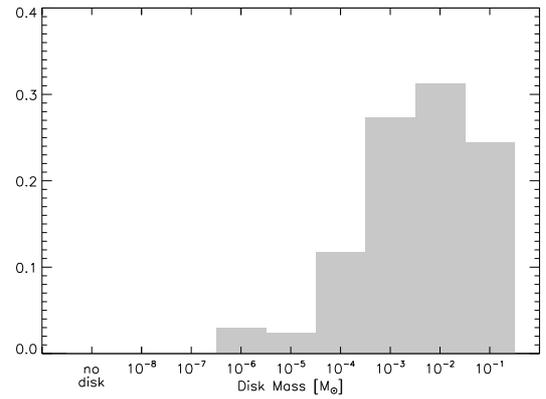}
\caption{\small Distribution of the disk mass [M$_\odot$]\label{fig:DMass}}
\end{figure}\\
Figure \ref{fig:DMass} shows the histogram of disk masses. We find that configurations with a disk of intermediate mass ($10^{-3}$ to $10^{-1}$\,M$_\odot$) are most probable. Nevertheless, we want to stress that \mbox{(sub-)}\,mm data with sufficient angular resolution are not available. Only such observations would give the total mass of circumstellar material by measuring the optically thin thermal dust emission. Without such data, any assumptions on the mass of the circumstellar material will be model dependent.\\
\begin{figure}[ht!]
\centering
\includegraphics[width=0.8\linewidth,keepaspectratio]{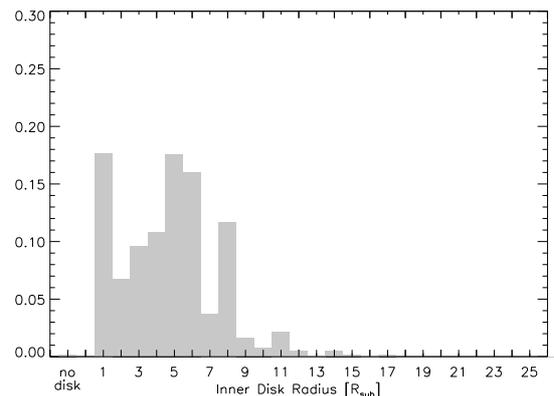}
\caption{\small Distribution of the Disk inner radius [$R_{\rm sub}$] model parameter. Values of -1 correspond to 
envelope\,-\,only models\label{fig:DInn}}
\end{figure}\\ 
Figure \ref{fig:DInn} shows the corresponding inner radii of the proposed disk structure, in units of the thermal dust 
sublimation radius. It is interesting to note that the most probable inner radii are larger than the dust destruction radius. This result is roughly consistent with the relation between the total luminosity and the source size as established by \citet{2007prpl.conf..539M}. However, the radius of the best\,-\,fitting model (Sect. \ref{sec:comparisson} table \ref{tab:3003929}) exceeds the dust sublimation radius by a factor of two, suggesting ongoing inner disk dispersal by photo\,-\,evaporation and\,/\,or stellar winds.
\subsection{Object models from the SED Fitting}
As a result of the SED fitting, we obtained 4 slightly different model configurations with various inclination angles which resemble the measured SED best (see table \ref{tab:SED-Data}). In table \ref{tab:3003929}, we summarized the most important structural parameters. A more detailed overview can be found in table \ref{longtable2} in the appendix.\vspace{5pt}\\
All of these model configurations consider a disk only; hence, no additional circumstellar material is needed to account for the near and mid\,-\,infrared fluxes. The central object is in every case quite massive. According to \citet{2005A&A...436.1049M}, the objects can be classified as early B stars. The disks themselves are massive as well, yet the distance of the inner rim to the central object differs considerably. Since the inner radius is directly related to the spatial distribution of the dust and, therefore, to the spatial intensity distribution, we should be able to distinguish the different models when comparing the synthetic and measured visibility curves. We explicitly do not specify an outer scale for the disk, since our data are not capable of confining the size of the disk in a model\,-\,independent way.
\begin{table}[ht!]
\caption{\small Parameters and derived quantities of the models. Note that these are disk\,-\,only models.\label{tab:3003929}}
\small
\begin{tabular*}{\linewidth}{ c | c  c  c  c }
\hline
\multirow{2}{*}{parameter}&\multicolumn{4}{c}{Grid Model Number}\\
 & 3003929 & 3005903 & 3008813 & 3012934 \\\hline
stellar mass [M$_\odot$] & 15 & 16 & 16 & 14\\
stellar temperature & \multirow{2}{*}{31300} & \multirow{2}{*}{32300} & \multirow{2}{*}{32200} & \multirow{2}{*}{30500}\\
$[$K$]$&\\
disk mass [M$_\odot$]& 9\,$\times10^{-2}$ & 4\,$\times10^{-2}$ & 2\,$\times10^{-2}$ & 6\,$\times10^{-2}$\\
disk inner radius & \multirow{2}{*}{26 (2.2)} & \multirow{2}{*}{84 (6.4)} & \multirow{2}{*}{99 (7.7)} & \multirow{2}{*}{55 (5.1)} \\
$[$AU$]$ ([$R_{\rm sub}^{\space}$])&\\
disk accretion rate & \multirow{2}{*}{7.8\,$\times10^{-8}$} & \multirow{2}{*}{3.0\,$\times10^{-6}$} & \multirow{2}{*}{2.2\,$\times10^{-7}$} & \multirow{2}{*}{4.9\,$\times10^{-7}$}\\
$[$M$_\odot$/yr$]$&\\
Luminosity & \multirow{2}{*}{20.7\,$\times10^{3}$} & \multirow{2}{*}{25.9\,$\times10^{3}$} & \multirow{2}{*}{25.0\,$\times10^{3}$} & \multirow{2}{*}{17.7\,$\times10^{3}$}\\
$[$L$_{\odot}]$ &\\
\hline
\end{tabular*}   
\end{table}
\subsection{General discussion of the visibilities}\label{sec:Gen_Visis}
\begin{figure}[ht!]
\centering
\includegraphics[width=\linewidth,keepaspectratio]{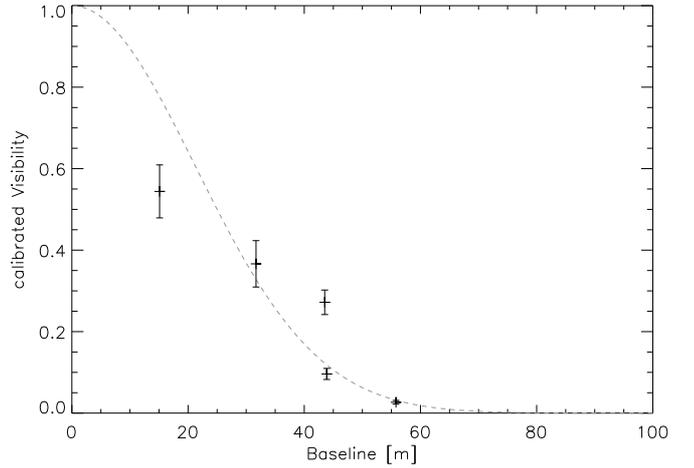}
\caption{\small Fit of a 1D Gaussian intensity distribution to the visibility amplitudes at 10.5\,$\mu$m. \label{fig:Gaussfit}}
\end{figure}

Before actually comparing the synthetic to the measured visibility curves, we will discuss a few more general conclusions which can be drawn from the overall shape of the measured curves themselves. The most intriguing fact is that although the baseline lengths for the 2005-06-24 and 2006-05-18 measurements are almost equal, the respective visibility amplitudes are clearly distinct, even within the depicted 3\,$\sigma$ errors (see Fig. \ref{pic:vis-curves}). The respective projected angles (141\degree\, and 48\degree) differ by approximately 90\degree. This is a clear indication for a non\,-\,rotationally symmetric intensity distribution, i.e. the mid\,-\,infrared emission is dominated by a flattened structure, e.g. a disk. This interpretation of the visibility curves is consistent with the disk\,-\,only scenario proposed by the statistical analysis of the SED fitting.\vspace{5pt}\\
To give an estimate of the size of the source, we fitted a Gaussian intensity profile\footnote{\tt $exp\big(-\frac{\pi^2}{4ln(2)}\frac{FWHM}{u}\big)$ ($FWHM\,[rad]$,$u=\frac{\lambda}{B}$)} to the visibility amplitudes at 8, 10.5 and 13\,$\mu$m (see Fig. \ref{fig:Gaussfit}, fit at $\lambda$=10.5\,$\mu$m). The corresponding full width at half maximum were determined as 33, 38 and 40\,mas (corresponding to 69, 80 and 84\,AU, if a distance of 2.1\,kpc is assumed), respectively. Note, though, that since obviously a Gaussian profile is a bad fit to the data, these numbers should be rather regarded as rough estimates.
\\ 
For instance, if one assumes an extended source of 8\,-\,13\,$\mu$m radiation, such as a halo or hot dust in an ultra compact H{\sc ii} region, one would expect to measure rather small visibility amplitudes. These sources typically have sizes larger than $5x10^{15}$\,cm (half\,-intensity radius, \citep{1990ApJ...354..247C}, corresponding to $\approx 160$\,mas at 2.1\,kpc), which would imply much lower visibilities at these baselines due to over\,-\,resolving the target (the respective FWHM in Fig. \ref{fig:Gaussfit} would be narrower by approximately a factor of 4 to 5). This is in accordance with the results found in the previous section, namely that the most probable solution for the SED fitting is an evolved, disk\,-\,only system without a dust envelope. The argumentation holds as well for every source of N\,-\,band emission being much larger than the scales found above. It applies, for example, also to a system consisting only of a central star plus flared circumstellar disk, seen edge\,-\,on. There, the direct view onto the hot inner rim of the disk is blocked, and we just receive more diluted emission on larger scales. Given our assumption that the projected intensity distribution of the KWO is not rotationally symmetric, which excludes a face\,-\,on disk, we would expect some intermediate inclination angle. Indeed, the ten best-fitting models are inclined by 57 to 81$^\circ$ (where the resulting configuration still allows a direct line of sight to parts of the central disk edges (see Fig. \ref{pic:model}).
\subsection{Comparison between modeled and measured visibility curves} \label{sec:comparisson}
Here, we discuss the comparison between the modeled synthetic visibility curves and the measured ones (see Figs. \ref{pic:comparison} and \ref{pic:comparison2}). Apparently, there is quite a discrepancy between the synthetic and the measured spectrally dispersed visibility curves. Nevertheless, there are clear differences in the qualitative shape of the modeled visibility curves, which enable us to exclude some of the models which otherwise fit the SED of the KWO almost equally well. Hence, this approach offers the ability to reduce the ambiguities arising from SED\,-\,only fitting.\vspace{5pt}\\
The difference between measurement and model can have several origins. First, the flux measurements used with the Online SED model fitter were incomplete. Lacking  submm/mm data, we were not able to constrain the flux from the cold dust in the outer parts of the disk. This relates directly to the ambiguity in the results of the Online SED model fitter (see Fig. \ref{pic:SEDFit}: The fitted SEDs differ considerably in this wavelength region). In addition, the modeling still suffers from many imperfections. For example, the model used here does not account for a puffed\,-\,up inner rim (\citet{2001A&A...371..186N}, \citet{2001ApJ...560..957D}), yet most of the flux in the N\,-\,band stems from regions of the disk which would be affected in dimension and shape if this were implemented. A change of the geometric shape of this region will affect the shape of the synthetic visibility curves. The dust model used by the RT simulation code is also not well adapted to this disk\,-\,only situation, since it assumes a dust composition that resembles the typical composition of the halo, not including bigger dust particles which are likely present in an evolved disk. Additionally, a bigger variety of dust particles will lead to a smoothing of the dust destruction zone. Smaller dust grains in this region reach their evaporation temperature at larger distances from the central object than bigger ones. Hence, the inner radius of the disk as a function of height correlates with the size of the dust grain constituents, leading to a more curved, rather than flat, hot inner rim (\citet{2007ApJ...661..374T}, \citet{2005A&A...438..899I}).\vspace{5pt}\\
Proper error handling is difficult to achieve when comparing the synthetic curves with the measurements. A conventional approach would be a global $\chi^2$ measure, including deviations on all 5 baselines. We found that many models fit all the measurements except the 55.8\,m baseline, where they fail to reproduce the rather low visibility amplitude. A conventional $\chi^2$ measure, on the other hand, would overweight these measurements due to their small absolute errors. Yet, we found that the relative error of all measurements is almost constant. Hence, instead of the absolute error, we used the square of the relative error as a weight in the error analysis. As a result, each curve is almost equally significant. In order to justify our ansatz, we did the same analysis excluding the 55.8\,m baseline. We found that in this case, the two ways of handling the error deliver the same results, while they deviate partially when we included all measurements in the standard deviation estimation. In this case, the absolute error weighted result was biased towards the 55.8\,m baseline. Hence, we decided to use the square of the relative error as the weight for comparing our measurements to the model. The results of this analysis are summarized in table \ref{tab:Visifit}. For a better comparison, the inclination angle and the $\chi^2$ obtained from the SED fit are specified for each model in the first two columns. In column 4, the position angles (see section \ref{sec:RTmodel}) best\,-\,fitting the data and in column 5 the respective estimates of the deviation between the synthetic models and the data are given.
{\begin{table}[ht!]
\caption{\small Standard deviations of the SED fits and estimates of the deviation between modeled and measured visibility curves  \label{tab:Visifit}}
\small
\begin{tabular*}{\linewidth}{ c | c  c  c  c }
\hline
Grid model&Inclin. & SED Fit & Pos. & Visibility\\
number & angle [\degree] & $\chi^2$ & angle [\degree]&  $\chi^2$ \\\hline
3003929                 & 76 & 141 & 16  & 73\\\hline
3005903                 & 81 & 189 & 167 & 248\\\hline
\multirow{4}{*}{3008813}& 63 & 190 & 176 & 125\\
                        & 70 & 153 & 173 & 145\\
                        & 76 & 123 & 171 & 155\\
                        & 81 & 151 & 171 & 210\\\hline
\multirow{4}{*}{3012934}& 57 & 194 & 84  & 260\\
                        & 63 & 199 & 89  & 284\\
                        & 70 & 189 & 92  & 317\\
                        & 76 & 188 & 92  & 338\\\hline
\end{tabular*}   
\end{table}}\\
\noindent The best agreement between model and data is obtained with model 3003929, the deviation estimate being almost a factor of two smaller than the next best one. It is interesting to note, that model 3003929 is not the best SED fit, although the respective values of the standard deviation of the SED fit for model 3003929 and the best fit are comparable. Model 3005903 and the four different variations of model 3012934 agree least with the data.  Hence, we exclude them from the further discussion. Note that the four realizations of model 3008813 differ mainly in the inclination angle. The comparison between these four models and the data shows a trend toward smaller inclination angles. Yet, due to the small differences between the four realizations, the ones with inclination angles of 63\degree, 70\degree and 76\degree\, can be considered to agree equally well with the data. A comprehensive overview of the parameters determining model 3003929 (as well as the other models) can be found in Table \ref{longtable2} in the appendix.\vspace{5pt}\\
Although model 3003929 agrees best with the data, there are obvious deviations between the synthetic and the measured visibility curves. To begin with, the synthetic visibility amplitude for the 55.8\,m baseline is higher than the measured one. A possible reason for this was given above: we consider this deviation to be systematic. There is good agreement for the other baselines, yet we find a qualitative discrepancy. Model 3003929 shows increased visibility amplitudes for the 8 to 9\,$\mu$m regime. We think that the source of this behavior is directly related to how the RT code handles the inner rim. This rather narrow, hot region, from which emission in this wavelength regime emanates, is not as resolved in comparison to the other models due to the small inner disk radius of 3003929. This leads to higher visibility amplitudes. If one assumes that, in reality, a puffed\,-\,up inner rim exists, the spatial region of the disk emitting in the spectral range between 8 and 9\,$\mu$m might be considerably more extended, leading to the lower visibility amplitudes we see in our measurements.
\section{Conclusions}
In this paper, we present VLTI observations of the Kleinmann\,-\,Wright object with baselines ranging between 15\,m and 56\,m. Our target has been resolved for all baselines, with average visibility amplitudes between 0.01 and 0.69. In order to analyze the interferometric data, we performed SED fitting using the web tool of \citet{2007ApJS..169..328R}. The resulting information, together with a  Monte Carlo radiative transfer simulation code developed by \citet{2003ApJ...598.1079W}, was then used to create synthetic visibility curves. We compared these synthetic curves to the measured visibilities, allowing us to substantially narrow down the range of suitable solutions obtained by mere SED fitting.\vspace{5pt}\\
While massive YSOs are usually thought to be associated with a significant circumstellar envelope which often dominates their infrared appearance, KWO does not show indications for a strong envelope component, and hence is probably more advanced in its evolution towards the main sequence. This is in accordance with \citet{1998A&A...332..999P}, who proposed that the classification as an embedded MYSO might be misleading, since it could be behind M17 SW but not actually embedded in it. We found that the silicate absorption feature is missing in all visibility curves. This indicates that the silicate absorption feature, which is clearly visible in the total flux spectrum, is mainly due to foreground structures not associated with the KWO. In this geometry the correlated and uncorrelated fluxes are equally affected; hence, the visibilities don't show any pronounced silicate features. In addition, our statistical interpretation of the SED fitting also indicates the absence of a dusty envelope. We find that the visibility curves in general can only be interpreted using an  intensity distribution deviating from spherical symmetry, which hints at a circumstellar\,-\,disk scenario. Finally, we deduce that a flared disk, inclined at 76\degree\, and with a mass of $\approx$0.1 M$_\odot$ and an inner rim radius of 26\,AU, reproduces the combination of SED fitting and interferometric data best. To obtain better constraints for the dimensions of the disk and the amount of circumstellar material in general, high\,-\,resolution submm\,/\,mm data will be needed.
\bibliographystyle{aa}
\bibliography{bibliography.bib}

\begin{thebibliography}{28}
\expandafter\ifx\csname natexlab\endcsname\relax\def\natexlab#1{#1}\fi

\bibitem[{{Chini} {et~al.}(2004){Chini}, {Hoffmeister}, {K{\"a}mpgen},
  {Kimeswenger}, {Nielbock}, \& {Siebenmorgen}}]{2004A&A...427..849C}
{Chini}, R., {Hoffmeister}, V.~H., {K{\"a}mpgen}, K., {et~al.} 2004, \aap, 427,
  849

\bibitem[{{Churchwell} {et~al.}(1990){Churchwell}, {Wolfire}, \&
  {Wood}}]{1990ApJ...354..247C}
{Churchwell}, E., {Wolfire}, M.~G., \& {Wood}, D.~O.~S. 1990, \apj, 354, 247

\bibitem[{{de Wit} {et~al.}(2010){de Wit}, {Hoare}, {Oudmaijer}, \&
  {Lumsden}}]{2010A&A...515A..45D}
{de Wit}, W.~J., {Hoare}, M.~G., {Oudmaijer}, R.~D., \& {Lumsden}, S.~L. 2010,
  \aap, 515, A45+

\bibitem[{{Dullemond} {et~al.}(2001){Dullemond}, {Dominik}, \&
  {Natta}}]{2001ApJ...560..957D}
{Dullemond}, C.~P., {Dominik}, C., \& {Natta}, A. 2001, \apj, 560, 957

\bibitem[{{Henning} {et~al.}(1984){Henning}, {Friedemann}, {Guertler}, \&
  {Dorschner}}]{1984AN....305...67H}
{Henning}, T., {Friedemann}, C., {Guertler}, J., \& {Dorschner}, J. 1984,
  Astronomische Nachrichten, 305, 67

\bibitem[{{Hoffmeister} {et~al.}(2008){Hoffmeister}, {Chini}, {Scheyda},
  {Schulze}, {Watermann}, {N{\"u}rnberger}, \& {Vogt}}]{2008ApJ...686..310H}
{Hoffmeister}, V.~H., {Chini}, R., {Scheyda}, C.~M., {et~al.} 2008, \apj, 686,
  310

\bibitem[{{Isella} \& {Natta}(2005)}]{2005A&A...438..899I}
{Isella}, A. \& {Natta}, A. 2005, \aap, 438, 899

\bibitem[{{Jaffe}(2004)}]{2004SPIE.5491..715J}
{Jaffe}, W.~J. 2004, in Presented at the Society of Photo-Optical
  Instrumentation Engineers (SPIE) Conference, Vol. 5491, New Frontiers in
  Stellar Interferometry, Proceedings of SPIE Volume 5491. Edited by Wesley A.
  Traub. Bellingham, WA: The International Society for Optical Engineering,
  2004., p.715, ed. W.~A. {Traub}, 715--+

\bibitem[{{Kim} {et~al.}(1994){Kim}, {Martin}, \&
  {Hendry}}]{1994ApJ...422..164K}
{Kim}, S.-H., {Martin}, P.~G., \& {Hendry}, P.~D. 1994, \apj, 422, 164

\bibitem[{{Kleinmann} \& {Wright}(1973)}]{1973ApJ...185L.131K}
{Kleinmann}, D.~E. \& {Wright}, E.~L. 1973, \apjl, 185, L131+

\bibitem[{{Laor} \& {Draine}(1993)}]{1993ApJ...402..441L}
{Laor}, A. \& {Draine}, B.~T. 1993, \apj, 402, 441

\bibitem[{{Leinert} {et~al.}(2003){Leinert}, {Graser}, {Waters}, {Perrin},
  {Jaffe}, {Lopez}, {Przygodda}, {Chesneau}, {Schuller}, {Glazenborg-Kluttig},
  {Laun}, {Ligori}, {Meisner}, {Wagner}, {Bakker}, {Cotton}, {de Jong},
  {Mathar}, {Neumann}, \& {Storz}}]{2003SPIE.4838..893L}
{Leinert}, C., {Graser}, U., {Waters}, L.~B.~F.~M., {et~al.} 2003, in
  Interferometry for Optical Astronomy II. Proc.~SPIE, Vol.~4838, ed. W.~A.
  {Traub}, 893--904

\bibitem[{{Leinert} {et~al.}(2004){Leinert}, {van Boekel}, {Waters},
  {Chesneau}, {Malbet}, {K{\"o}hler}, {Jaffe}, {Ratzka}, {Dutrey}, {Preibisch},
  {Graser}, {Bakker}, {Chagnon}, {Cotton}, {Dominik}, {Dullemond},
  {Glazenborg-Kluttig}, {Glindemann}, {Henning}, {Hofmann}, {de Jong},
  {Lenzen}, {Ligori}, {Lopez}, {Meisner}, {Morel}, {Paresce}, {Pel},
  {Percheron}, {Perrin}, {Przygodda}, {Richichi}, {Sch{\"o}ller}, {Schuller},
  {Stecklum}, {van den Ancker}, {von der L{\"u}he}, \&
  {Weigelt}}]{2004A&A...423..537L}
{Leinert}, C., {van Boekel}, R., {Waters}, L.~B.~F.~M., {et~al.} 2004, \aap,
  423, 537

\bibitem[{{Linz} {et~al.}(2009){Linz}, {Henning}, {Feldt}, {Pascucci}, {van
  Boekel}, {Men'shchikov}, {Stecklum}, {Chesneau}, {Ratzka}, {Quanz},
  {Leinert}, {Waters}, \& {Zinnecker}}]{2009A&A...505..655L}
{Linz}, H., {Henning}, T., {Feldt}, M., {et~al.} 2009, \aap, 505, 655

\bibitem[{{Linz} {et~al.}(2008){Linz}, {Stecklum}, {Follert}, {Henning}, {van
  Boekel}, {Men'shchikov}, {Pascucci}, \& {Feldt}}]{2008JPhCS.131a2024L}
{Linz}, H., {Stecklum}, B., {Follert}, R., {et~al.} 2008, Journal of Physics
  Conference Series, 131, 012024

\bibitem[{{Martins} {et~al.}(2005){Martins}, {Schaerer}, \&
  {Hillier}}]{2005A&A...436.1049M}
{Martins}, F., {Schaerer}, D., \& {Hillier}, D.~J. 2005, \aap, 436, 1049

\bibitem[{{Millan-Gabet} {et~al.}(2007){Millan-Gabet}, {Malbet}, {Akeson},
  {Leinert}, {Monnier}, \& {Waters}}]{2007prpl.conf..539M}
{Millan-Gabet}, R., {Malbet}, F., {Akeson}, R., {et~al.} 2007, in Protostars
  and Planets V, ed. B.~{Reipurth}, D.~{Jewitt}, \& K.~{Keil}, 539--554

\bibitem[{{Natta} {et~al.}(2001){Natta}, {Prusti}, {Neri}, {Wooden}, {Grinin},
  \& {Mannings}}]{2001A&A...371..186N}
{Natta}, A., {Prusti}, T., {Neri}, R., {et~al.} 2001, \aap, 371, 186

\bibitem[{{Pinte} {et~al.}(2008){Pinte}, {Padgett}, {Menard}, {Stapelfeldt},
  {Schneider}, {Olofsson}, {Panic}, {Augereau}, {Duchene}, {Krist},
  {Pontoppidan}, {Perrin}, {Grady}, {Kessler-Silacci}, {van Dishoeck},
  {Lommen}, {Silverstone}, {Hines}, {Wolf}, {Blake}, {Henning}, \&
  {Stecklum}}]{2008arXiv0808.0619P}
{Pinte}, C., {Padgett}, D.~L., {Menard}, F., {et~al.} 2008, ArXiv e-prints, 808

\bibitem[{{Porter} {et~al.}(1998){Porter}, {Drew}, \&
  {Lumsden}}]{1998A&A...332..999P}
{Porter}, J.~M., {Drew}, J.~E., \& {Lumsden}, S.~L. 1998, \aap, 332, 999

\bibitem[{{Reimann} {et~al.}(2000){Reimann}, {Linz}, {Wagner}, {Relke},
  {Kaeufl}, {Dietzsch}, {Sperl}, \& {Hron}}]{2000SPIE.4008.1132R}
{Reimann}, H., {Linz}, H., {Wagner}, R., {et~al.} 2000, in Society of
  Photo-Optical Instrumentation Engineers (SPIE) Conference Series, Vol. 4008,
  Society of Photo-Optical Instrumentation Engineers (SPIE) Conference Series,
  ed. {M.~Iye \& A.~F.~Moorwood}, 1132--1143

\bibitem[{{Robitaille}(2008)}]{2008ASPC..387..290R}
{Robitaille}, T.~P. 2008, in Astronomical Society of the Pacific Conference
  Series, Vol. 387, Massive Star Formation: Observations Confront Theory, ed.
  H.~{Beuther}, H.~{Linz}, \& T.~{Henning}, 290--+

\bibitem[{{Robitaille} {et~al.}(2007){Robitaille}, {Whitney}, {Indebetouw}, \&
  {Wood}}]{2007ApJS..169..328R}
{Robitaille}, T.~P., {Whitney}, B.~A., {Indebetouw}, R., \& {Wood}, K. 2007,
  \apjs, 169

\bibitem[{{Robitaille} {et~al.}(2006){Robitaille}, {Whitney}, {Indebetouw},
  {Wood}, \& {Denzmore}}]{2006ApJS..167..256R}
{Robitaille}, T.~P., {Whitney}, B.~A., {Indebetouw}, R., {Wood}, K., \&
  {Denzmore}, P. 2006, \apjs, 167, 256

\bibitem[{{Siebenmorgen} {et~al.}(2004){Siebenmorgen}, {Kr{\"u}gel}, \&
  {Spoon}}]{2004A&A...414..123S}
{Siebenmorgen}, R., {Kr{\"u}gel}, E., \& {Spoon}, H.~W.~W. 2004, \aap, 414, 123

\bibitem[{{Tannirkulam} {et~al.}(2007){Tannirkulam}, {Harries}, \&
  {Monnier}}]{2007ApJ...661..374T}
{Tannirkulam}, A., {Harries}, T.~J., \& {Monnier}, J.~D. 2007, \apj, 661, 374

\bibitem[{{Whitney} {et~al.}(2003){Whitney}, {Wood}, {Bjorkman}, \&
  {Cohen}}]{2003ApJ...598.1079W}
{Whitney}, B.~A., {Wood}, K., {Bjorkman}, J.~E., \& {Cohen}, M. 2003, \apj,
  598, 1079

\bibitem[{{Zinnecker} \& {Yorke}(2007)}]{2007ARA&A..45..481Z}
{Zinnecker}, H. \& {Yorke}, H.~W. 2007, \araa, 45, 481

\end{thebibliography}
{\begin{figure*}[h!]
\centering
\begin{minipage}[t!]{.48\linewidth}
\includegraphics[width=\linewidth,keepaspectratio]{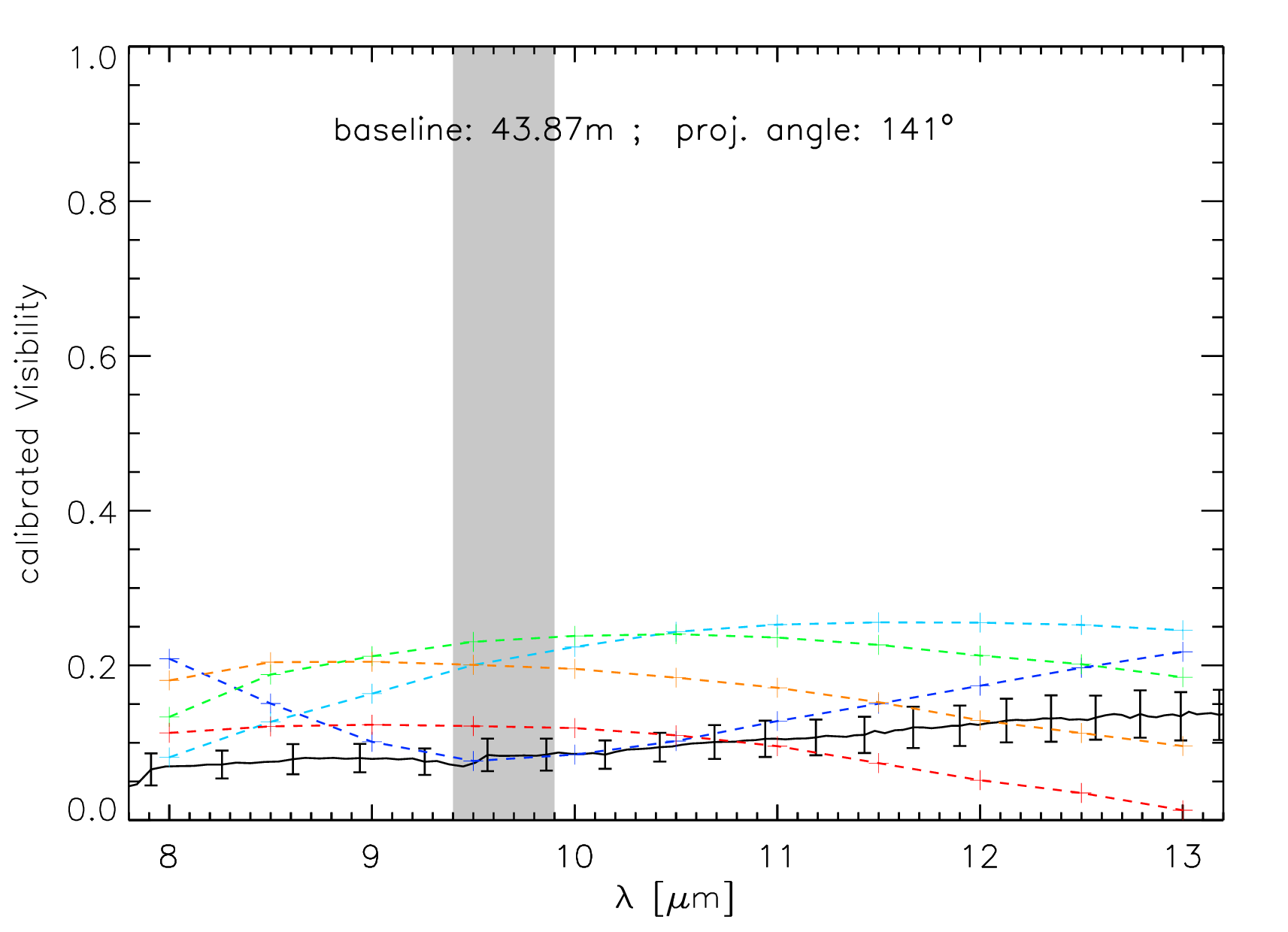}
\end{minipage}
\hspace{.02\linewidth}
\begin{minipage}[t!]{.48\linewidth}
\includegraphics[width=\linewidth,keepaspectratio]{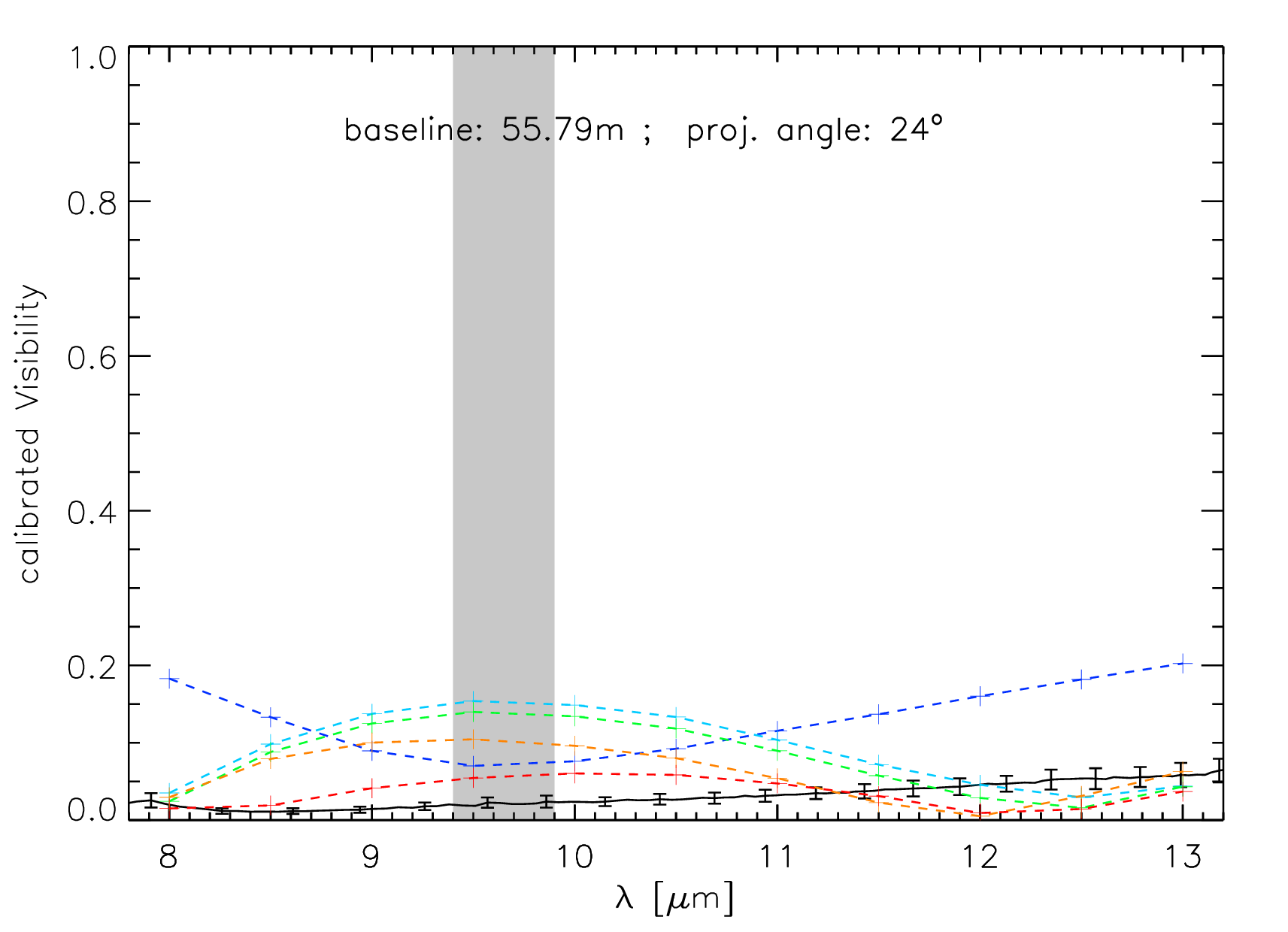}
\end{minipage}
\vspace{1cm}\\
\begin{minipage}[t!]{.48\linewidth}
\includegraphics[width=\linewidth,keepaspectratio]{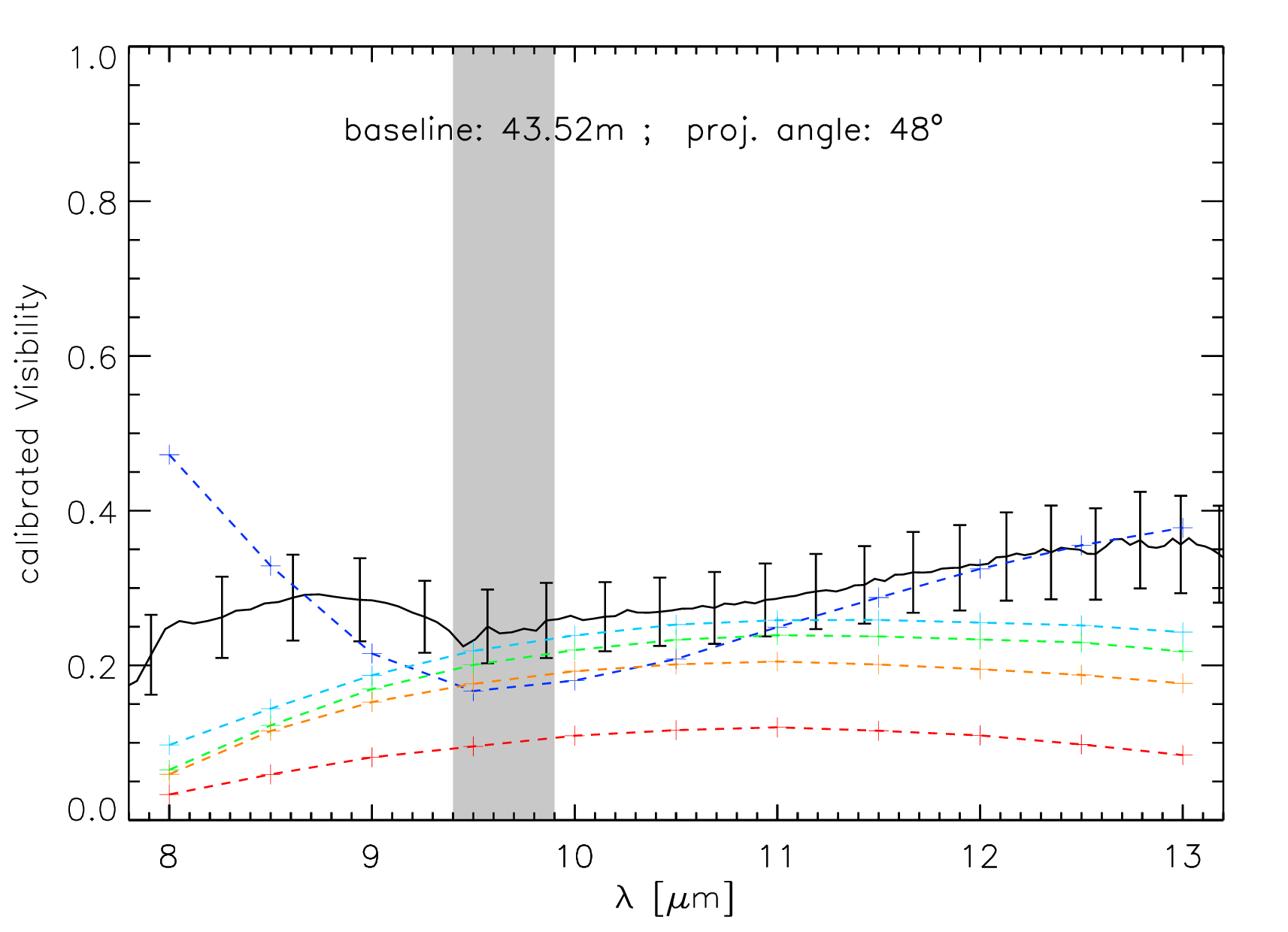}
\end{minipage}
\hspace{.02\linewidth}
\begin{minipage}[t!]{.48\linewidth}
\includegraphics[width=\linewidth,keepaspectratio]{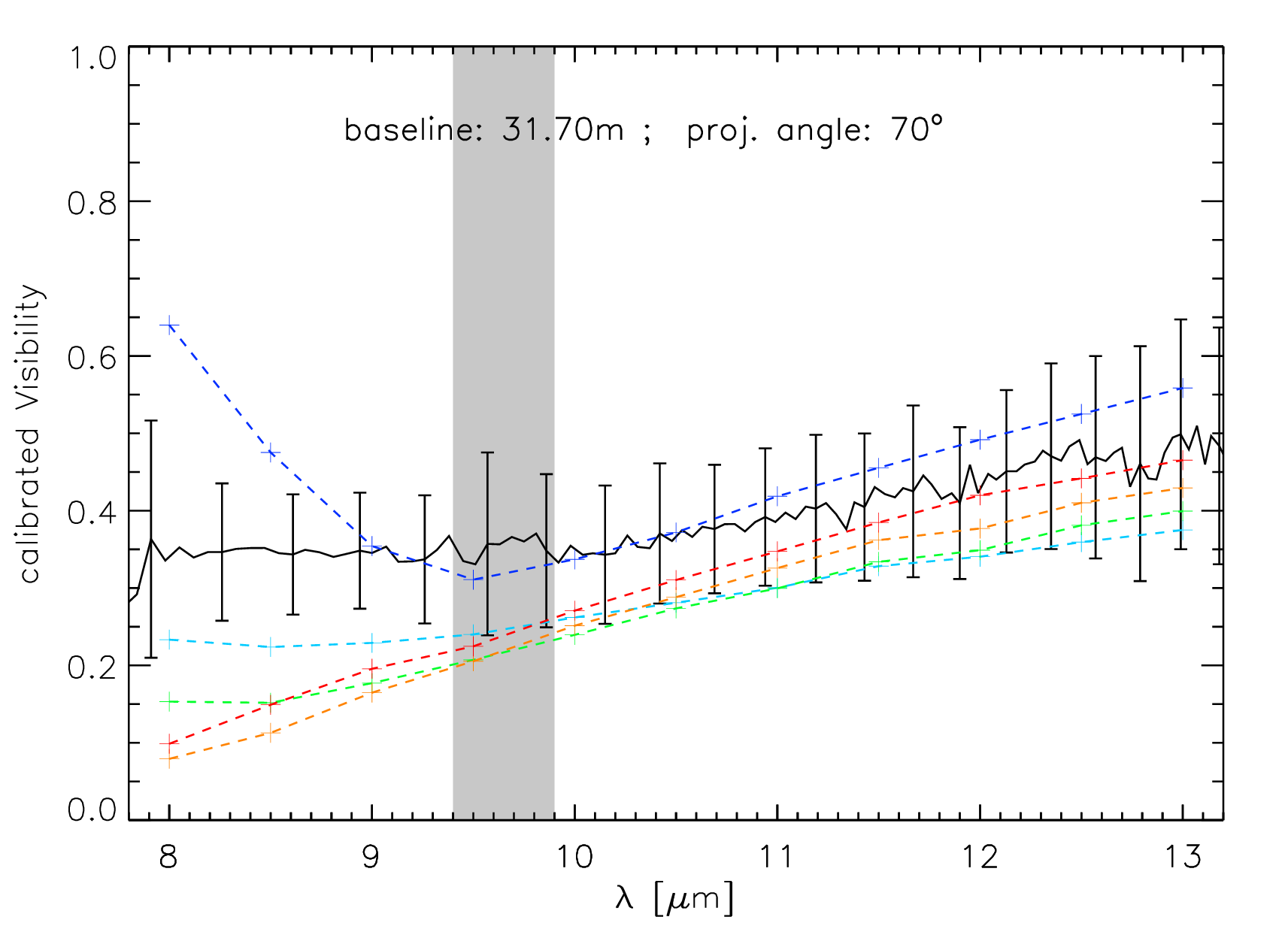}
\end{minipage}
\vspace{1cm}\\
\begin{minipage}[t!]{.48\linewidth}
\includegraphics[width=\linewidth,keepaspectratio]{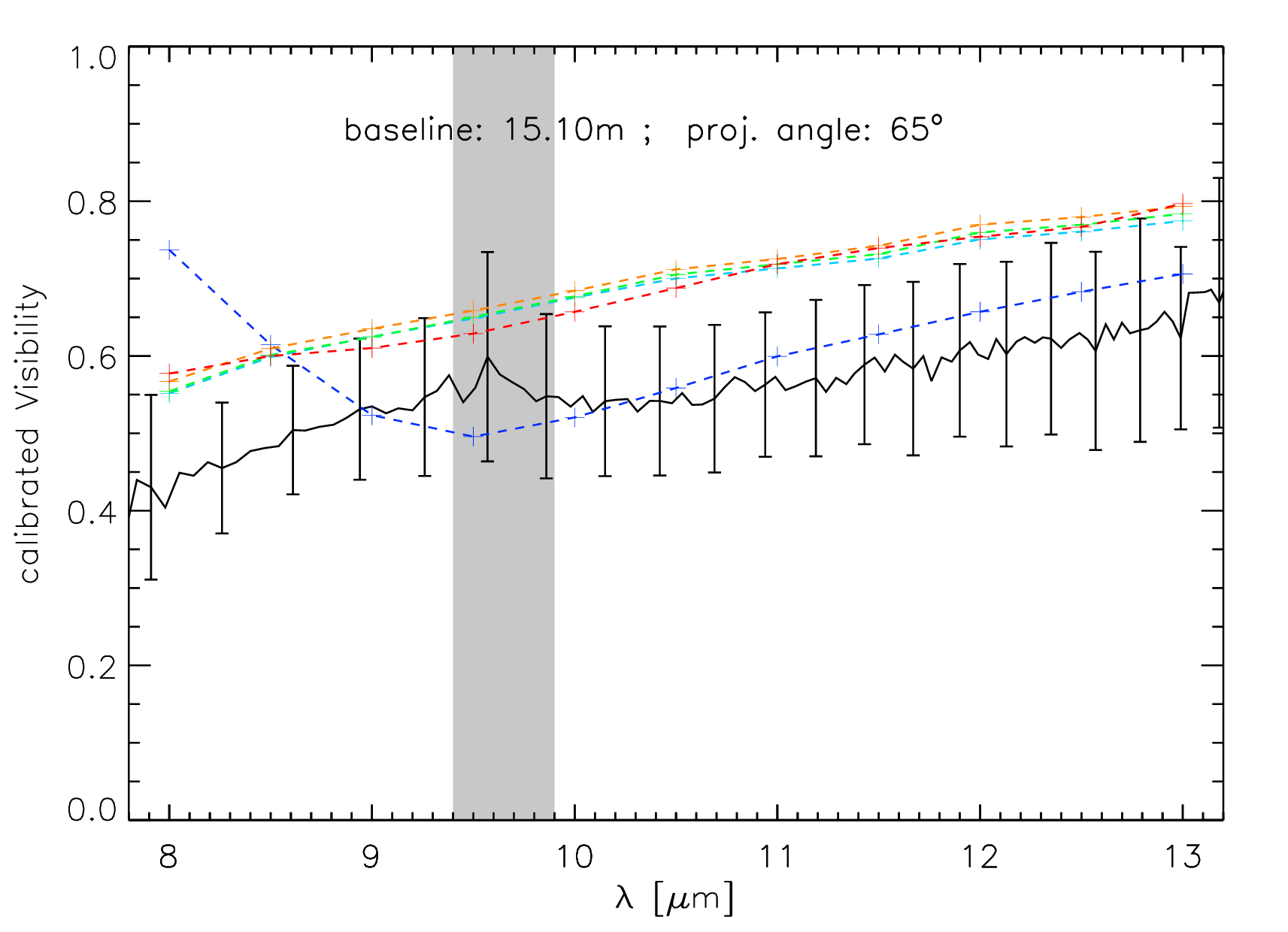}
\end{minipage}
\hspace{.02\linewidth}
\begin{minipage}[[t!]{.48\linewidth}
\centering
\includegraphics[width=\linewidth,keepaspectratio]{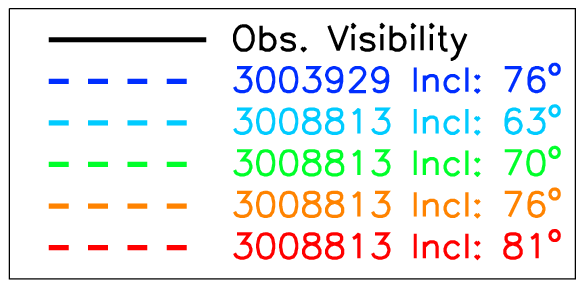}
\end{minipage}
\caption{Comparison of the observed and synthetic visibilities; each figure represents one baseline configuration. The error bars are 3$\sigma$.}
\label{pic:comparison}
\end{figure*}}\clearpage
{
\begin{figure*}[h!]
\centering
\begin{minipage}[t!]{.48\linewidth}
\includegraphics[width=\linewidth,keepaspectratio]{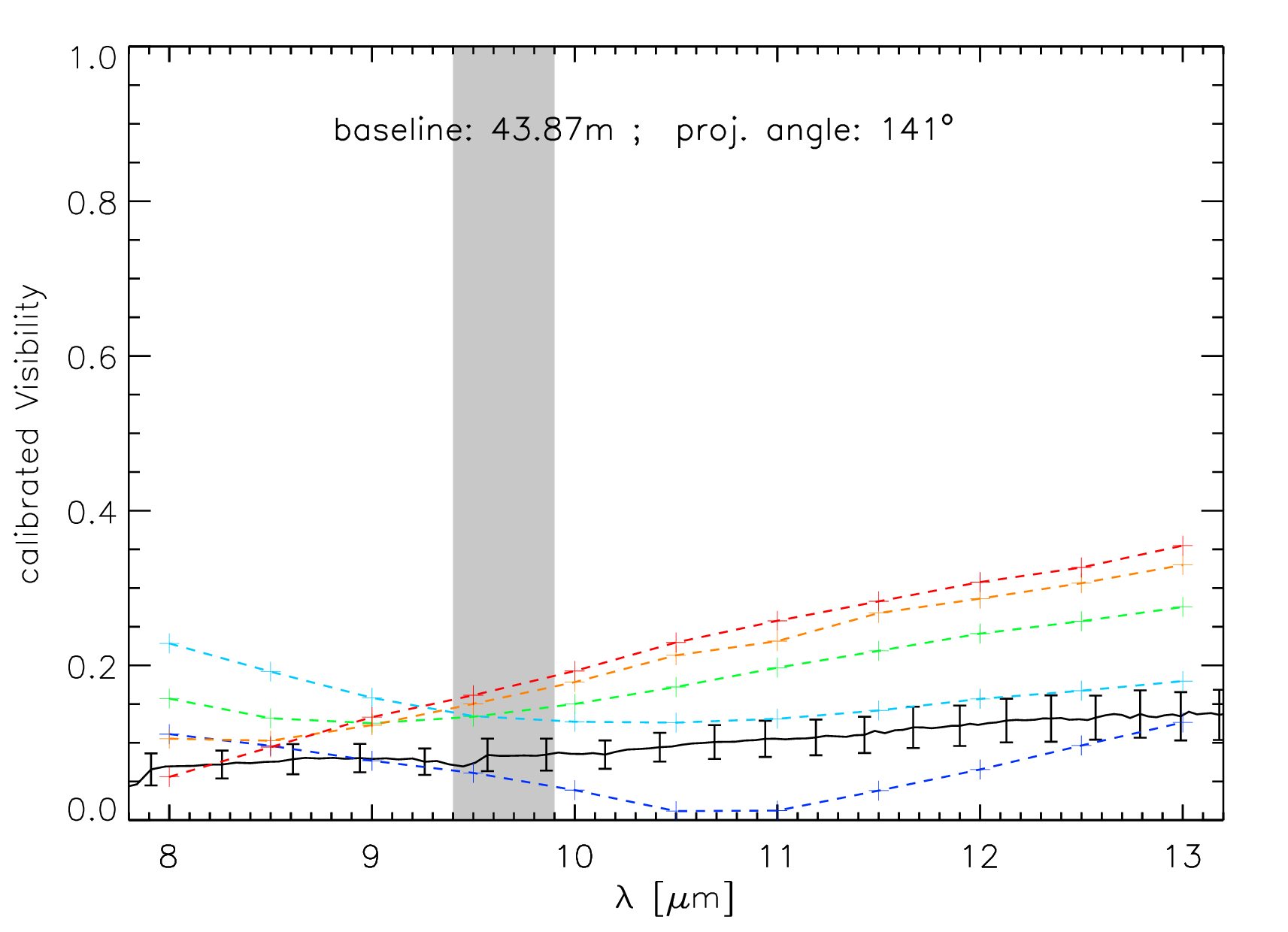}
\end{minipage}
\hspace{.02\linewidth}
\begin{minipage}[t!]{.48\linewidth}
\includegraphics[width=\linewidth,keepaspectratio]{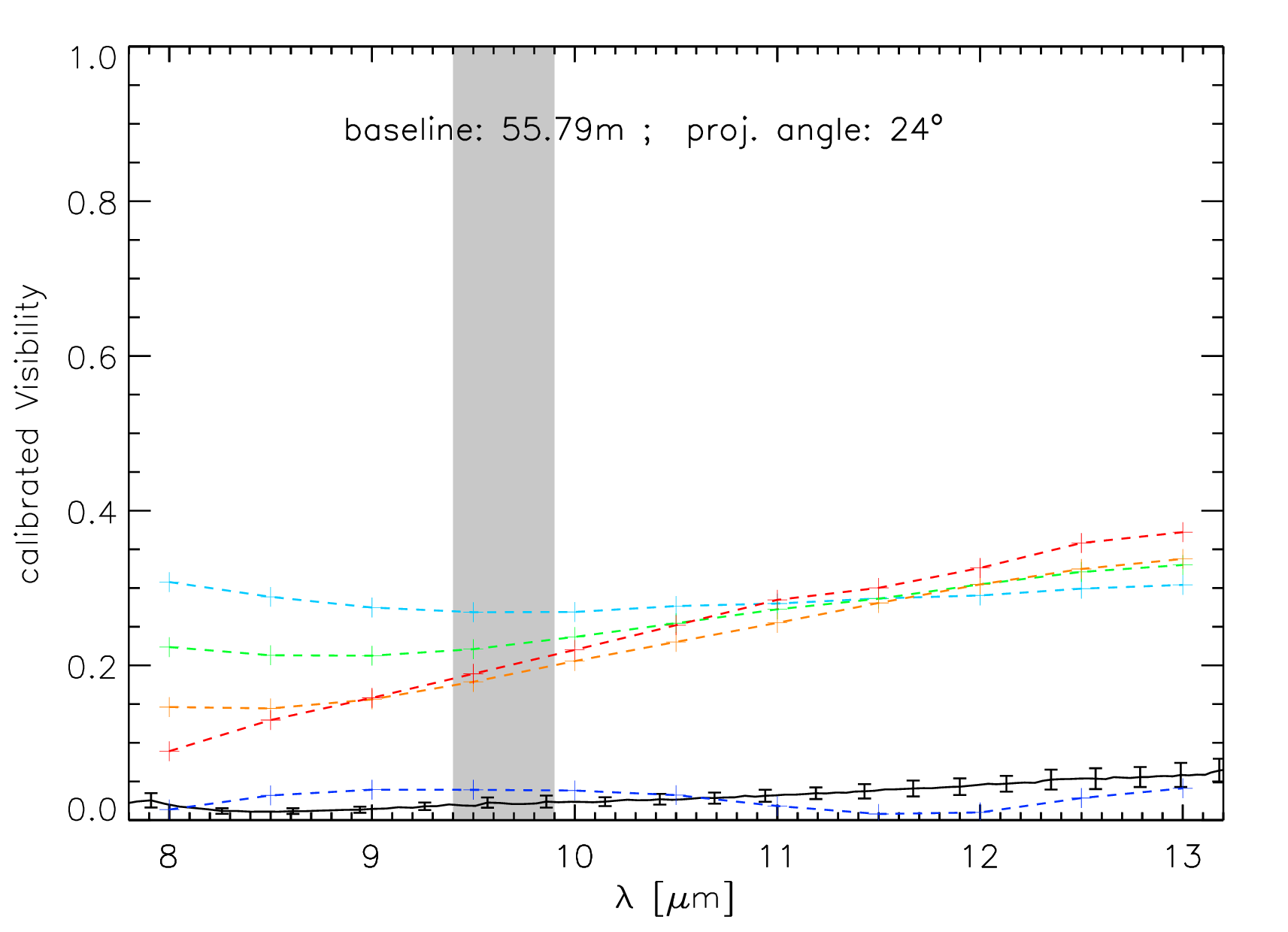}
\end{minipage}
\vspace{1cm}\\
\begin{minipage}[t!]{.48\linewidth}
\includegraphics[width=\linewidth,keepaspectratio]{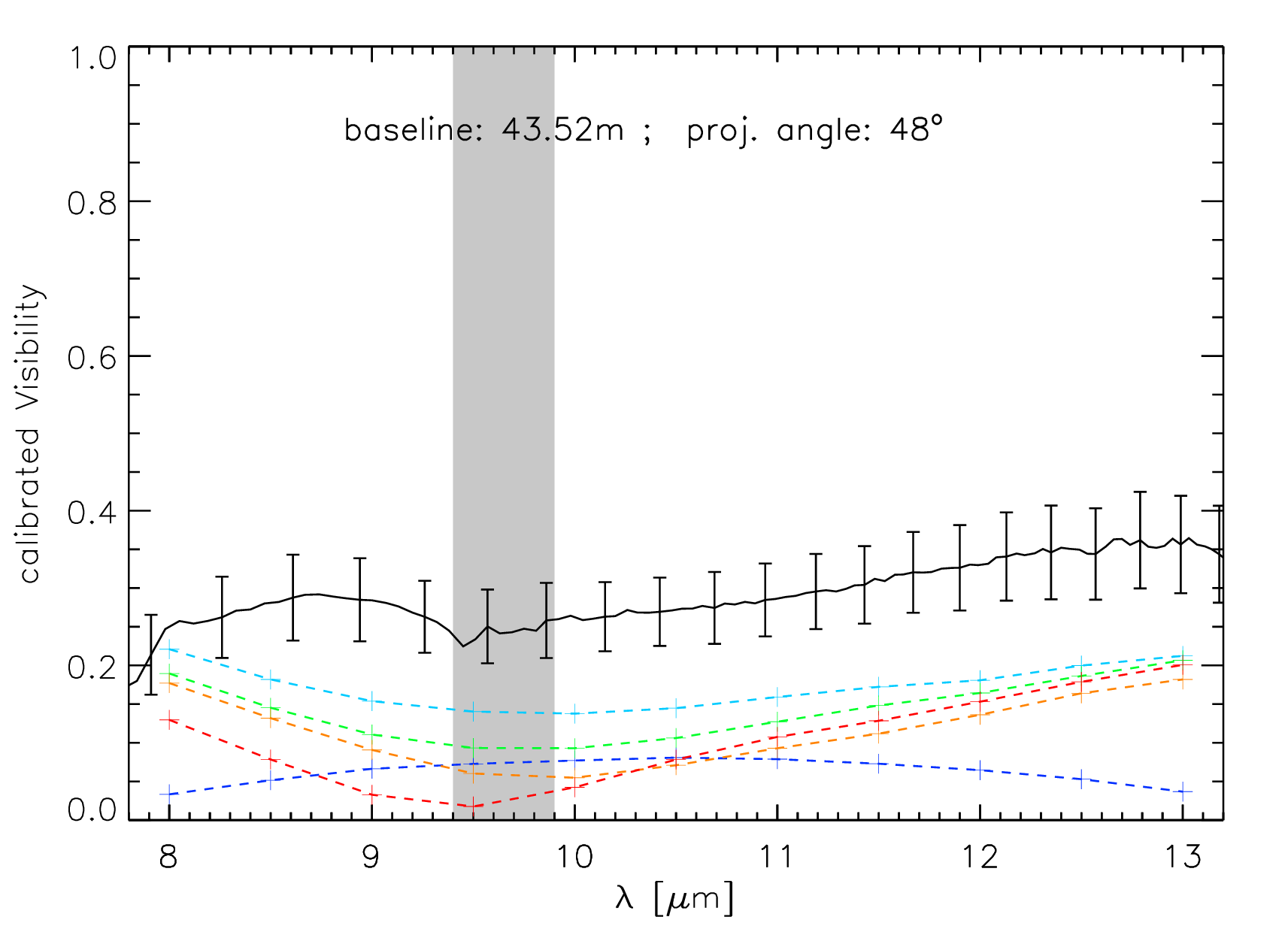}
\end{minipage}
\hspace{.02\linewidth}
\begin{minipage}[t!]{.48\linewidth}
\includegraphics[width=\linewidth,keepaspectratio]{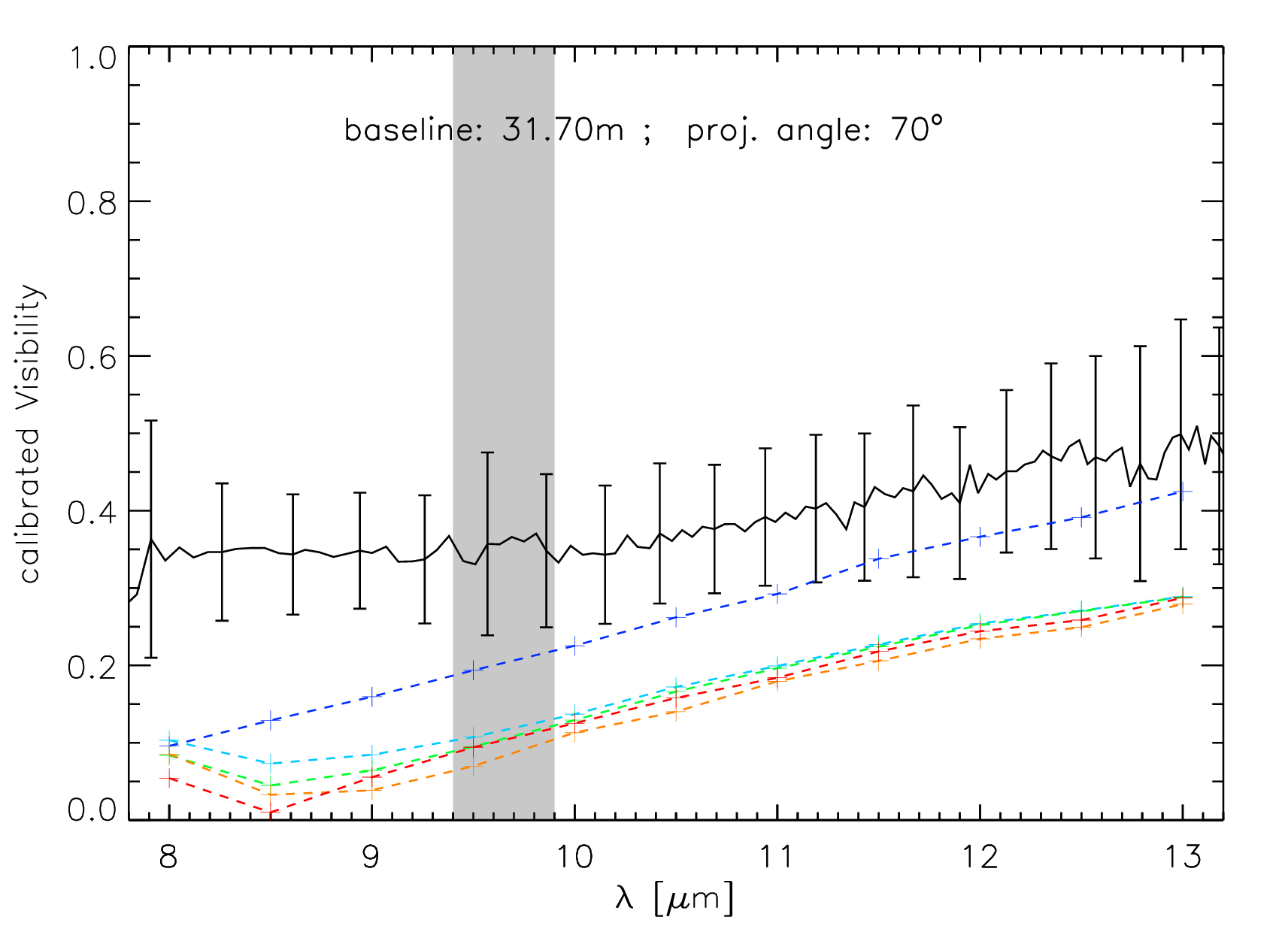}
\end{minipage}
\vspace{1cm}\\
\begin{minipage}[t!]{.48\linewidth}
\includegraphics[width=\linewidth,keepaspectratio]{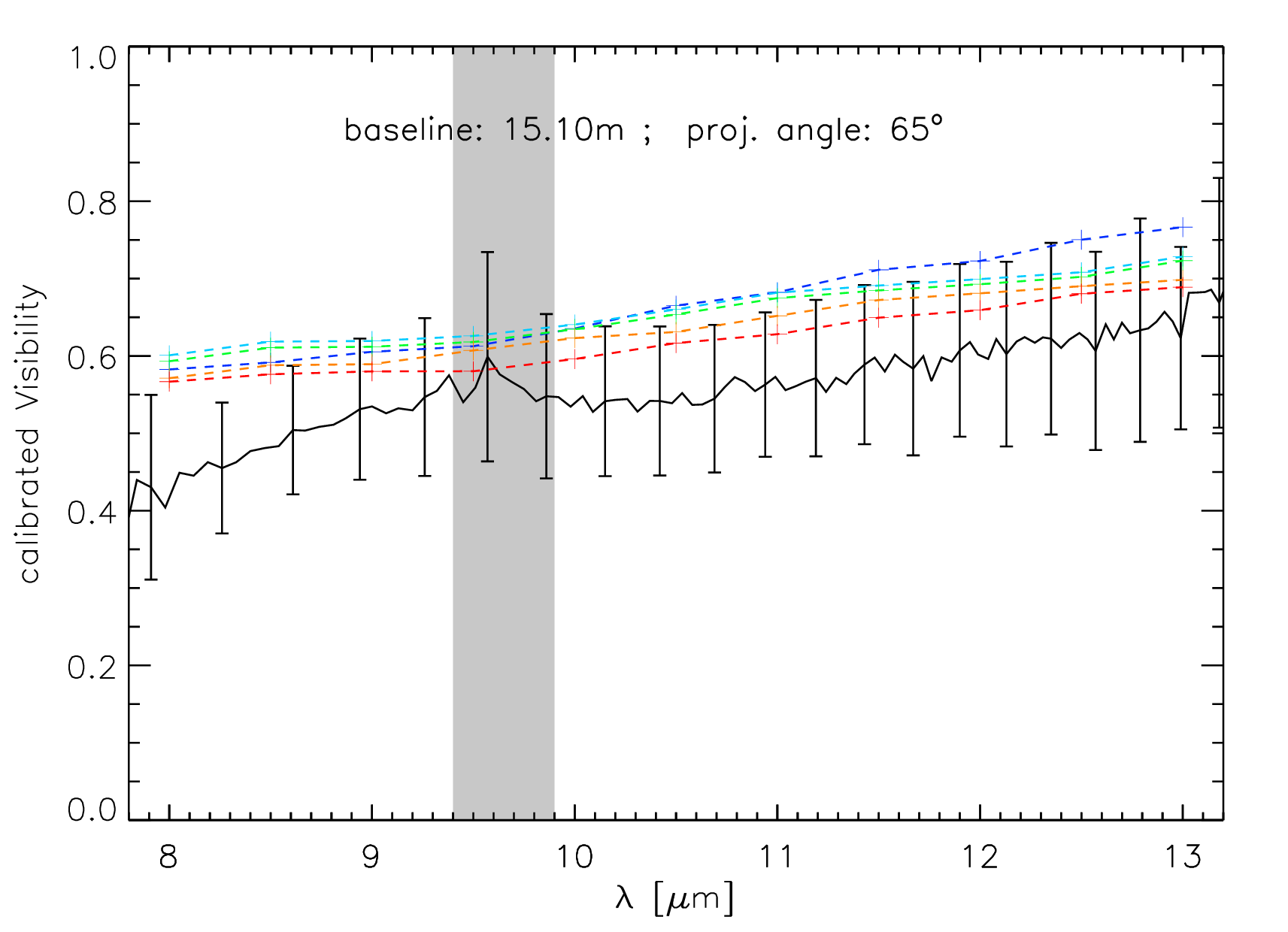}
\end{minipage}
\hspace{.02\linewidth}
\begin{minipage}[[t!]{.48\linewidth}
\centering
\includegraphics[width=\linewidth,keepaspectratio]{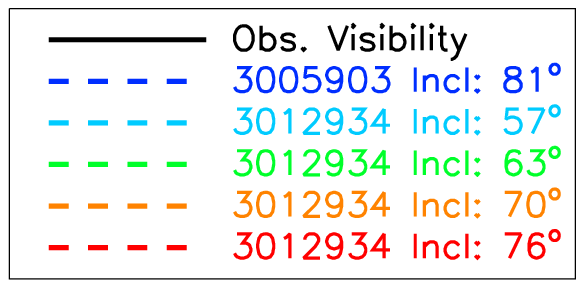}
\end{minipage}
\caption{Comparison of the observed and synthetic visibilities; each figure represents one baseline configuration. The error bars are 3$\sigma$.}
\label{pic:comparison2}
\end{figure*}}\clearpage
\begin{table}[t!]
\caption{Summary of model parameters. For details on the parameter definitions see \citet{2006ApJS..167..256R}.\label{longtable2}}
\begin{tabular}{|c|c|c|c|c|c|c|c|c|c|c|}\hline
\multicolumn{1}{|c|}{}&\multicolumn{1}{c|}{3003929}&\multicolumn{1}{c|}{3005903}&\multicolumn{4}{c|}{3008813}&\multicolumn{4}{c|}{3012934}\\\hline
Inclination&\multirow{2}{*}{76}&\multirow{2}{*}{81}&\multirow{2}{*}{63}&\multirow{2}{*}{70}&\multirow{2}{*}{76}&\multirow{2}{*}{81}&\multirow{2}{*}{57}&\multirow{2}{*}{63}&\multirow{2}{*}{70}&\multirow{2}{*}{76}\\
Angle[\degree]&&&&&&&&&&\\
SED fitting&\multirow{2}{*}{141}&\multirow{2}{*}{189}&\multirow{2}{*}{190}&\multirow{2}{*}{153}&\multirow{2}{*}{123}&\multirow{2}{*}{151}&\multirow{2}{*}{194}&\multirow{2}{*}{199}&\multirow{2}{*}{189}&\multirow{2}{*}{188}\\
$\chi^2$&&&&&&&&&&\\
distance&\multirow{2}{*}{1905}&\multirow{2}{*}{1995}&\multirow{2}{*}{2291}&\multirow{2}{*}{2291}&\multirow{2}{*}{2291}&\multirow{2}{*}{2188}&\multirow{2}{*}{2188}&\multirow{2}{*}{2188}&\multirow{2}{*}{2089}&\multirow{2}{*}{1995}\\
 $[pc]$&&&&&&&&&&\\
$A_\nu$&\multirow{2}{*}{134}&\multirow{2}{*}{587}&\multirow{2}{*}{19}&\multirow{2}{*}{18}&\multirow{2}{*}{17}&\multirow{2}{*}{94}&\multirow{2}{*}{19}&\multirow{2}{*}{18}&\multirow{2}{*}{19}&\multirow{2}{*}{27}\\
(total)&&&&&&&&&&\\
$A_\nu$&\multirow{2}{*}{13}&\multirow{2}{*}{12}&\multirow{2}{*}{19}&\multirow{2}{*}{18}&\multirow{2}{*}{17}&\multirow{2}{*}{12}&\multirow{2}{*}{19}&\multirow{2}{*}{18}&\multirow{2}{*}{19}&\multirow{2}{*}{16}\\
(interstellar)&&&&&&&&&&\\
stellar&\multirow{2}{*}{4.9}&\multirow{2}{*}{5.1}&\multirow{2}{*}{5.1}&\multirow{2}{*}{5.1}&\multirow{2}{*}{5.1}&\multirow{2}{*}{5.1}&\multirow{2}{*}{4.8}&\multirow{2}{*}{4.8}&\multirow{2}{*}{4.8}&\multirow{2}{*}{4.8}\\
radius [R$_\odot$]&&&&&&&&&&\\
stellar&\multirow{2}{*}{15}&\multirow{2}{*}{16}&\multirow{2}{*}{16}&\multirow{2}{*}{16}&\multirow{2}{*}{16}&\multirow{2}{*}{16}&\multirow{2}{*}{14}&\multirow{2}{*}{14}&\multirow{2}{*}{14}&\multirow{2}{*}{14}\\
mass [M$_\odot$]&&&&&&&&&&\\
stellar&\multirow{2}{*}{31000}&\multirow{2}{*}{32000}&\multirow{2}{*}{32000}&\multirow{2}{*}{32000}&\multirow{2}{*}{32000}&\multirow{2}{*}{32000}&\multirow{2}{*}{31000}&\multirow{2}{*}{31000}&\multirow{2}{*}{31000}&\multirow{2}{*}{31000}\\
temperature [K]&&&&&&&&&&\\
Luminosity&\multirow{2}{*}{20700}&\multirow{2}{*}{25900}&\multirow{2}{*}{25000}&\multirow{2}{*}{25000}&\multirow{2}{*}{25000}&\multirow{2}{*}{25000}&\multirow{2}{*}{17700}&\multirow{2}{*}{17700}&\multirow{2}{*}{17700}&\multirow{2}{*}{17700}\\
(total,[L$_\odot$])&&&&&&&&&&\\
envelope accr.&\multirow{2}{*}{0}&\multirow{2}{*}{0}&\multirow{2}{*}{0}&\multirow{2}{*}{0}&\multirow{2}{*}{0}&\multirow{2}{*}{0}&\multirow{2}{*}{0}&\multirow{2}{*}{0}&\multirow{2}{*}{0}&\multirow{2}{*}{0}\\
rate [M$_\odot$/yr] &&&&&&&&&&\\
disk mass&\multirow{2}{*}{9x10$^{-2}$}&\multirow{2}{*}{4x10$^{-2}$}&\multirow{2}{*}{2x10$^{-2}$}&\multirow{2}{*}{2x10$^{-2}$}&\multirow{2}{*}{2x10$^{-2}$}&\multirow{2}{*}{2x10$^{-2}$}&\multirow{2}{*}{6x10$^{-2}$}&\multirow{2}{*}{6x10$^{-2}$}&\multirow{2}{*}{6x10$^{-2}$}&\multirow{2}{*}{6x10$^{-2}$}\\
$[M_\odot]$&&&&&&&&&&\\
disk inner&\multirow{2}{*}{2}&\multirow{2}{*}{6}&\multirow{2}{*}{8}&\multirow{2}{*}{8}&\multirow{2}{*}{8}&\multirow{2}{*}{8}&\multirow{2}{*}{5}&\multirow{2}{*}{5}&\multirow{2}{*}{5}&\multirow{2}{*}{5}\\
radius [R$_{sub}$]&&&&&&&&&&\\
disk inner&\multirow{2}{*}{26}&\multirow{2}{*}{84}&\multirow{2}{*}{99}&\multirow{2}{*}{99}&\multirow{2}{*}{99}&\multirow{2}{*}{99}&\multirow{2}{*}{55}&\multirow{2}{*}{55}&\multirow{2}{*}{55}&\multirow{2}{*}{55}\\
radius [AU]&&&&&&&&&&\\
disk accr.&\multirow{2}{*}{8x10$^{-08}$}&\multirow{2}{*}{3x10$^{-6}$}&\multirow{2}{*}{2x10$^{-7}$}&\multirow{2}{*}{2x10$^{-7}$}&\multirow{2}{*}{2x10$^{-7}$}&\multirow{2}{*}{2x10$^{-7}$}&\multirow{2}{*}{5x10$^{-7}$}&\multirow{2}{*}{5x10$^{-7}$}&\multirow{2}{*}{5x10$^{-7}$}&\multirow{2}{*}{5x10$^{-7}$}\\
rate [M$_\odot$/yr]&&&&&&&&&&\\
disk flaring&\multirow{2}{*}{1.15}&\multirow{2}{*}{1.04}&\multirow{2}{*}{1.10}&\multirow{2}{*}{1.10}&\multirow{2}{*}{1.10}&\multirow{2}{*}{1.10}&\multirow{2}{*}{1.10}&\multirow{2}{*}{1.10}&\multirow{2}{*}{1.10}&\multirow{2}{*}{1.10}\\
parameter $\beta$&&&&&&&&&&\\
\hline
\end{tabular}
\end{table}
\noindent Note that the total extinction given in the table is the sum of the interstellar and intrinsic extinction. The intrinsic extinction for disk\,-\,like structures is given for the direct line of sight; hence, it depends considerably on the inclination and flaring of the disk. If the total and interstellar extinction are equal on the other hand, this indicates a direct line of sight to the central stellar source.\\
\citet{2006ApJS..167..256R} state that the disk structure is given by:
\begin{displaymath}
\rho(\varpi,z) = \rho_0 \Big[1-\sqrt{\frac{R_*}{\varpi}}\Big]\Big(\frac{R_*}{\varpi}\Big)^{\alpha}exp\Big(-\frac{1}{2}\big[\frac{z}{h}\big]^2\Big) \nonumber
\end{displaymath}
where $h \propto \varpi^{\beta}$ and $\alpha=\beta+1$.
\end{document}